# Irregular Metamaterial Networks

*Metamaterials derived from irregular networks promise remarkable functionality but require a new type of network science.*

Thomas P. Wytock[a,b], Chiara Daraio[c], Heinrich M. Jaeger[d,e], Christopher A. Schuh[f], Lorenzo Valdevit[g], Vincenzo Vitelli[d,e,h] & Adilson E. Motter[a,b,i,j]

## Abstract

Metamaterials can achieve exceptional functionality through careful engineering of their mesoscale structure. Although appropriately introduced irregularities can be advantageous, current approaches largely conform to regular structures to preserve tractability. Here, we contend that network theory, enriched with geometry and physics, provides a natural framework for designing metamaterials with controlled irregularities at relevant scales, thereby enabling the discovery of new property-enhancing structures. We examine how this augmented network theory can facilitate the creation of irregular metamaterials with enhanced or novel properties and how metamaterial research, in turn, is opening new directions in network science. Supported by machine learning and advanced self-assembly, the emerging field of irregular metamaterial networks is poised to transform inverse design and scalable manufacturing of novel materials.

---

[a] Department of Physics and Astronomy, Northwestern University, Evanston, IL, USA.
[b] Center for Network Dynamics, Northwestern University, Evanston, IL, USA.
[c] Division of Engineering and Applied Science, California Institute of Technology, Pasadena, CA, USA.
[d] Department of Physics, University of Chicago, Chicago, IL, USA.
[e] James Franck Institute, University of Chicago, Chicago, IL, USA.
[f] Department of Materials Science and Engineering, Northwestern University, Evanston, IL, USA.
[g] Department of Materials Science and Engineering, University of California, Irvine, CA, USA.
[h] Leinweber Center for Theoretical Physics, University of Chicago, Chicago, IL, USA.
[i] Department of Engineering Sciences and Applied Mathematics, Northwestern University, Evanston, IL, USA.
[j] Northwestern Institute on Complex Systems, Northwestern University, Evanston, IL, USA.

## Introduction

The rise of metamaterials research over the past 30 years has established that the effective material properties need not be limited to those of ordinary materials synthesized at the atomic or molecular scale[1]. Metamaterials are architected to gain properties from their mesoscopic structure in addition to their microscopic composition. Examples have included metamaterials exhibiting exceptional specific strength[2,3], negative refractive index[4], negative Poisson's ratio[5,6], and sound confinement properties[7,8], with applications ranging from lightweight aerospace platforms, optical cloaking, and mechanical allostery to acoustic superlensing. Traditionally, metamaterials consist of regular patterns, formed by repeating structures whose form and orientation are designed with precision. Recent research shows that a new generation of metamaterials with desired properties is possible by departing from the stringent requirement of regularity (Fig. 1). The key to unlocking this potential is network science.

The study of network systems, whose origin is often traced to Euler, gained attention in physics with Maxwell's 1864 study of networks of forces[9]. Through the second half of the 20$^{th}$ century, physical scientists placed emphasis on the regular lattice networks of crystal-like systems while mathematicians, motivated by the structure of communication networks, developed the important field of random graph theory[10]. Over the past 25 years, it has become clear that real physical, biological, and social networks tend to fall in between these two extremes: they are irregular but also highly structured and thus by no means random[11,12]. The study of the structure and dynamics of such network systems—ranging from whole-brain connectomes[13–15] and continental-scale power grids[16] to global-scale social media[17,18]—has led to the emergence of network science as a cross-disciplinary field of research that interacts synergistically with emerging AI technologies and has proved effective, for instance, in supporting COVID-19 pandemic studies. More recently, an emerging direction in network theory is to consider the physical properties of network nodes and edges, exploring how volume exclusion influences the structure[19,20] and how the structure can be designed to exhibit desired dynamics[21] and material properties[22].

Network concepts had been previously used, often implicitly, in both the analysis and design of materials. The analysis of networks of forces gave rise to the field of topological mechanics[23], which studies how the structure of elements like trusses produces robust mechanical behaviors. Rigidity, in particular, has been formulated in this context as a vector percolation problem and was used to identify rigid and floppy domains in 2D random truss lattices[24], though efficient algorithms do not yet exist for general network structures. The dynamical matrix[25], which contains the network connectivity information, is a major tool for linear response analysis in topological mechanics previously applied to truss[26] and granular[27] materials. A pressing question is how to extend these ideas to capture a broad range of nonlinear and large-deformation behaviors.

In parallel, the seminal work establishing the relationship between microscale network structure and macroscopic material behavior characterized the elastic properties of both naturally occurring[28,29] and conventionally synthesized materials[30,31]. This effort led to the discovery of distinct scaling relationships across deformation regimes and the development of materials selection charts[28,32] for the comprehensive assessment of cost versus performance. One outcome of this line of research was the realization that architected materials could achieve performance outside the property space occupied by existing materials[33]. The design of such materials has been facilitated by topology optimization[34], which is a computational approach that combines linear programming with finite-element analysis. Although not readily scalable to irregular metamaterials, this approach has successfully assisted in the optimization of regular and locally coupled materials[35,36].

This Perspective has the two complementary goals of showing (i) that irregular metamaterials can exhibit enhanced properties relative to regular ones and (ii) that a suitable *network* representation of such materials facilitates their description, analysis, prediction, and design. The term *irregular* is used to refer to systems that are not formed by repeated building blocks but that may still be highly organized (as is common in optimized structures), and thus it includes, but is not limited to, (disordered) materials with random deviations from regularity. We distinguish between topological, geometrical, and parametric irregularity, discuss how they are naturally accounted for in network representations, and present measures to quantify them. We explore two representative network models to illustrate and overcome the challenges associated with the generation of physical networks. This enables us to explore the relationship between physical networks and machine learning, and to identify fruitful ways to deploy them together. Having established the potential of irregular metamaterial networks, we review how additive manufacturing techniques can be used to fabricate the proposed materials and highlight outstanding problems and opportunities for future research.

## Metamaterial Networks

A recurrent claim in network science has been that network structure influences—or even determines—function. For example, the percolation threshold in random networks is determined by the moments of the distribution of degrees[12] (where degree is the number of interactions per node), which underlie the well-established vanishing of epidemic threshold in scale-free networks[37]. In the master stability formalism[12], the stability of network states is governed by the eigenvalues of the Laplacian matrix, which, in the context of synchronization, generally implies that synchronized behavior is more strongly supported in networks with a localized spectrum of non-identically zero eigenvalues[38]. Similarly, the structural bias by which nodes in a complex network are likely to be connected with nodes that are more highly connected than themselves[12] ensures that dynamical processes are disproportionately mediated by highly connected nodes, thereby shaping the speed and pathways of diffusion and influence[39].

Crucially, the connection between structure and function is believed to largely hold in materials, too[30,31]. For instance, the allotropes of carbon are all made up of the same atoms, and thus the dissimilar properties of, say, diamond and graphite must emerge from the network structure (i.e., topology, geometry, and nature of interactions). Conversely, the tetrahedral covalent network of diamond is also found in other ultra-hard materials with different elemental compositions, such as silicon carbide and carbon nitrides[40,41]. In the materials literature, the method of topology optimization formalizes the structure-function relationship through finite-element descriptions of the filled material volume[42,43]. This approach can program preferred deformation modes known as compliant mechanisms by employing optimization to determine where material elements should be deposited in the finite-element grid[44]. The architecture and nature of interactions in both ordinary materials and those constructed via topology optimization are constrained by the adjacency of chemical and physical bonds. This limits the overall network structures to being generally locally coupled, whether or not they are ordered as in the case of crystals, designed as in the case of topology optimized structures, or disordered as in amorphous solids. Moreover, the metamaterials literature has focused mainly on structures that are not only locally coupled but also largely regular, since these constraints facilitate design by keeping the search space more tractable, and local coupling restricts the extent of irregularity that can be explored.

Importantly, these limitations are not inherent to metamaterials, enabling the exploration of architectures that extend beyond locally coupled and regular designs, despite the prevailing emphasis on them thus far. Metamaterials with novel or enhanced properties can be realized by exploring more general networks, as this increases the space of possibilities and hence the opportunities to bring about new behavior. While many properties of interest (e.g., specific stiffness and strength, wave attenuation at specific frequencies) are achieved by a periodic design, recent research on neuromorphic metamaterials[45], nature-designed structures[46] (including those of protein, cartilage, and muscle[47]), damage-resistant topologies[48,49,50], and disordered auxetic systems[51] indeed shows that substantial gains are possible by departing from regular and/or locally coupled network structures. Further advances in irregular metamaterials have been inspired by naturally occurring disordered materials such as foam, glass, granular, and amorphous systems, as well as (irregular) polymers, biological tissues, and animal-built structures. A promising question for future research is: how to comprehensively design and analyze metamaterials by systematically exploring a wide range of irregular networks (through optimization)?

Figure 2 illustrates how irregular granular, truss, and plate materials can be described by the same network structure. The elements of the underlying structure can be associated with material constituents of increasing complexity, which interact through points of contact in granular systems (Fig. 2a), elastic beams in truss lattices (Fig. 2b), and surfaces in plate materials (Fig. 2c). Properties of the underlying network structure (Fig. 2d), such as percolation and eigenvalue spectrum, can inform material behavior such as rigidity[52] and floppy modes[23]. As the example shows, this network representation also has important limitations. First, the network was derived from the material in Fig. 2a and then used to generate the materials in Fig. 2b,c, as opposed to be having been generated *a priori* based on desired material properties. Second, the description in Fig. 2d, while flexible, does not uniquely define the material, allowing for the vastly different outcomes shown in the figure as well as others. Therefore, the network representation must be augmented with additional information to render a complete description of the material and to inform analysis and design.

**Beyond Irregular Topology**

Networks can be—and in other domains they generally are—irregular in their full scale[53] and may involve directed coupling[54]. A metamaterial designed with this structure would not have finite-size unit cells in the thermodynamic limit, may exhibit a divergent variance for the number of interactions per network node, and could potentially undergo new types of phase transitions due to non-reciprocal interactions[55]. Yet, current network theory cannot be immediately deployed to design such irregular metamaterials because, apart from interaction strengths, traditional network models describe only the *topology* of the network. Metamaterials are embedded in the physical space and, as such, also require modeling the *geometry* and *nature* of the network elements.

**Traditional network models**. Figure 3 illustrates some of the most widely employed models in network science[11,12]. These models are visually represented as dots (nodes) connected by lines (edges) and expressed as matrices $A = (A_{ij})$, where $A_{ij}$ is the strength of the edge from node $j$ to node $i$ (Fig. 3a-d). The power of such models lies in their ability to capture essential properties frequently observed in real networks—including short node-to-node distances[56], skewed distribution of connections per node[57], and community structure[58]—while facilitating the characterization of emerging phenomena for entire network ensembles. Generalizations beyond static pair-wise interactions of a single type have also been explored through models accounting for higher-order[59,60], multilayer[61,62], and temporal interactions[63,64] (Fig. 3e-g). Notwithstanding their usefulness, these models are essentially limited to describing a system's topology. While there are physically embedded networks, such as power grids[16], social proximity networks[65], pedestrian flow networks[66], and the Internet[67], that are meaningfully captured by variants of traditional models[53], we argue below that the network modeling of metamaterials requires further extensions of network theory itself.

**Types of irregularity**. Progress in *irregular* metamaterial networks requires a suitable definition and quantification of irregularity. Figure 4 shows different types of network irregularities that can be explored in the design of metamaterials. At the most fundamental level, we have *topological* irregularity, which concerns the form of the underlying graph with no reference to physical distances. This form of irregularity is important, for example, in determining percolation transitions[68]. Next, we have *geometrical* irregularity, in which the size, shape, and/or orientation of the network nodes and/or edges may vary unevenly in space. Finally, we have *parametric* irregularity, which concerns node and edge properties not captured by topology and geometry. These properties define the nature of the network elements and may include mass density, composition, damping coefficients, and elastic moduli, among others. By dividing the types of irregularity into these categories, we arrive at a natural conceptualization of irregular metamaterials that applies both to engineered and natural materials[69] (Fig. S1). In all cases, the irregularity can be localized (e.g., a defect), structured (as in the case of gradients and holes), or statistically uniform across the entire network (as in the case of uniformly random systems). The degree of irregularity will generally depend on the scale at which the metamaterial is observed. As for other material network properties, this scale dependence can be quantified using persistent homology analysis[70].

**Quantifying irregularity**. By definition, irregular metamaterials deviate from regular patterns formed by repeating structures, as reflected in the underlying network and associated parameters. Concrete measures to quantify this deviation include symmetry-, entropy-, fluctuation-, and information-based metrics.

*Network symmetries* can be defined in terms of the network's automorphism group[71] and its generalizations. This group is formed by all node permutations $i \to \pi(i)$ that leave the network unchanged. Structurally identical nodes are therefore related by a permutation and belong to the same orbit of the group (symmetries can be defined similarly for permutations among edges). The orbits form a partition of the network. It is immediate that regular network structures will have more symmetries than irregular ones, which in the most extreme cases may have no symmetry other than the identity. Network symmetries have been used to characterize network irregularity in other contexts[72,73] and hold promise in the study of metamaterials.

Entropy is a measure of disorder often used in dynamical contexts, but it can nevertheless quantify structural irregularity. In an ensemble of networks, each with an occurrence probability $P_n$, such *configurational entropy*[74] can be defined to be proportional to the Gibbs entropy $S = -k_B \Sigma_n P_n \ln P_n$, with $k_B$ the Boltzmann constant. An ensemble consisting of one or a few configurations, such as a lattice network, will have much lower entropy than an ensemble of random networks. The probabilities $P_n$ depend on the details of how the ensemble is defined[75,76].

For example, random networks generated from a given degree sequence define a microcanonical ensemble in which each network structure with the specified degrees has an equal probability. The Erdős–Rényi model (Fig. 3a), on the other hand, defines a canonical ensemble, in which all network structures are possible but with an occurrence probability proportional to a network version of the Boltzmann factor.

The *local volume-fraction variance* is another quantity that can be used to measure structural irregularity. This quantity is the variance of the fraction of the volume occupied by the material within a ball of radius $R$ as the ball's center is varied[77]. For $d$-dimensional materials, the volume fraction often scales as $R^{-(d+\xi)}$, where $\xi = 1$ for lattices, $0 < \xi \leq 1$ for a range of disordered hyperuniform structures, $\xi = 0$ for uniform randomness, and $-d < \xi < 0$ for anti-hyperuniform structures. This fluctuation-based characterization of irregularity directly captures long-range spatial correlations, which are of relevance to the design of metamaterials[78].

Shannon's *information entropy*[79,80] can be used to create yet another measure of irregularity. It can be applied to a realization of the material network covered with a fine grid in which grid boxes overlapping with the material are assigned 1, and empty boxes are assigned 0. Similar in form to the Gibbs entropy but very different in content, Shannon's entropy for this sequence of zeros and ones gives the amount of information needed to specify the structure of the system. Once again, a repetitive structure is expected to require less information than an irregular one.

While the measures above are presented to characterize irregularity in the material structure, they can be naturally extended to account for (possibly continuous) parameter dependences, such as the elastic modulus. The measures can thus quantify not only topological but also geometric and parametric irregularities, as well as combinations of them.

**Geometry-Endowed Network Models**

The first requirement for a metamaterial network is geometry. But therein lies the challenge: top-down direct attempts to incorporate geometry in existing generative models of irregular networks often lead to intersections and crossings of nodes and edges in the physical space. The prohibition of such overlaps can be referred to as the *volume exclusion constraint*. Another requirement is that *edge distribution must decay* with edge length. Indeed, due to volume exclusion and resource limitations, an edge between two nodes should become less likely if the nodes are sufficiently far apart. Such a property is featured in several models of physically embedded networks, most notably in the description of synaptic connectomes[81,82] (but also in spatial network models that are not constrained by volume exclusion[83-88]). In contrast with many other physically embedded network systems, such as microfluidics, brains, power grids, and computer chips, mechanical metamaterials are not static within the relevant time scales. Thus, geometry-endowed network models need to be developed to satisfy the volume exclusion constraint not only in a fixed equilibrium state, but also when stressed, strained, acoustically driven, heated, and/or damaged. An emerging body of research in the network science community has begun to grapple with the effects of volume exclusion[19,20], physical edge entanglement[89-91], and physically extended nodes[92,93]. A gallery of recent results in the field of physical networks is presented in **Box 1**. In what follows, we will explain the types of irregularity that the networks can exhibit and present two models illustrating the promise and peril of physical networks.

This line of research has two important frontiers: the first one concerns the elastic behavior of the nodes and edges so that volume exclusion can be respected under loading. The second involves accounting for the cost of physical edges as considered in recent work describing how the morphology of natural physical networks is driven by surface optimization[93] and as illustrated next.

**Endowing topological network models with geometry.** To appreciate the impact of geometry, we consider a simple geometric extension of the Erdős–Rényi model. In its standard formulation, the model is purely topological, where each pair of nodes is connected independently with fixed probability $p$. In our 3D geometric extension: (i) nodes are centered at randomly selected points in the unit cube, with no more than one node within each voxel of a $S^3$ grid; (ii) an edge is added between each pair of nodes $(i,j)$ with probability $\min\{\, p\, r_{ij}{}^{-\alpha},\ 1\}$ assuming a total of $N$ nodes and periodic boundary conditions, where $r_{ij}$ is the (Manhattan) distance between the nodes, $\alpha \geq 0$ is the decay exponent, and $p > 0$ is a constant; (iii) nodes are balls and edges are rods of radius $d \geq 0$. This model can be used to incorporate several previously considered concepts, like the edge probability decay with distance[85,88,94], volume exclusion[19,20], edge entanglement[89], and percolation theory[86,95], into a single construction.

Figure 5 summarizes some salient properties of this model. For $d = 0$, volume exclusion is automatically satisfied for any finite number of nodes, and the placement of each edge is independent of all others. However, even in this limit, there are surprises. For example, the average degree at the percolation transition—where a *giant* connected component forms—increases from 1 when $0 \leq \alpha \lesssim 2.5$ to larger than 1 as $\alpha \gtrsim 2.5$ when $N = 6250$ and $S = 50$ (Fig. 5b). Notwithstanding, the degree needed to obtain a *single* connected component with a fixed probability $e^{-1}$ remains fixed at $\frac{1}{2}\ln N$, which is the same threshold as in the (purely topological) Erdős–Rényi model (see Supplementary Information for additional model details). The latter result means that the decay parameter $\alpha$ naturally tunes the network to reach this connectivity transition with less material, since the typical edge length shortens as $\alpha$ increases (Fig. 5c). Moreover, edge/node crossings can occur as the system is subject to a non-affine straining and, in the example of Fig. 5d, the fraction of edge crossings grows linearly with the maximum strain. As soon as edge/node thickness is non-negligible, overlaps can occur with a frequency that grows linearly with $d$ even in the absence of any loading (Fig. 5e). Such volume exclusion violations can be addressed by allowing edges to have curvature and/or by fusing overlapping sections, which causes the edges to no longer be independent when $d > 0$. This model is representative of a broader ongoing initiative to develop physical network models that incorporate geometry into topological network designs (**Box 1**).

**Building networks from adjacency-constrained tiling.** The virtual growth model[96] is a recently conceived approach that automatically accommodates the physical constraints above. The model consists of building blocks that are tiled together randomly while satisfying certain compatibility rules (see Fig. 6). A key ingredient of the model is that each building block includes not only nodes and edge segments but also their embedding in the respective two- or three-dimensional space. Given a set of building blocks and compatibility rules, the model is entirely parameterized by the assigned proportion of building blocks of different types. The model's name derives from the generation process, which starts with one block, to which a second compatible block is attached, and so on, thereby mimicking nest design strategies adopted by insects using predominantly only local information[97]. As such, this is a bottom-up approach from which ensembles of emergent irregular networks can be generated systematically and physically realized via 3D printing (see Fig. 6e-f). The resulting complexity is intermediate between regular lattices (consisting of repetitions of a single building block) and random networks (which do not include repeated patterns).

A defining attribute of irregular metamaterials is the degree of correlations created by different arrangements of the same building blocks, as illustrated in Fig. 6b-d. This attribute impacts the percolation and physical properties of the material[98], and yet it is easy to show that it is not properly captured by widely used topological and geometric descriptors such as the Minkowski functionals[99]. These functionals, which in two dimensions comprise the Euler characteristic, perimeter, and area of the material structure, characterize size and shape "additively": each Minkowski functional of the whole material is a function of the fraction of each type of building block and material connections formed between adjacent building blocks. Because additive descriptors ignore correlations, holistic network-based descriptors are needed to capture the nonadditive emergent properties of metamaterials, as further discussed below.

It is important to compare the two types of network models just mentioned. The first treats nodes and edges as the fundamental building blocks. It enables long-range interactions but requires an explicit treatment of volume exclusion. The second, by contrast, adopts fundamental building blocks comprising nodes, edges, *and* empty space. This approach constrains long-range interactions but automatically enforces volume exclusion and can be realized through purely local, bottom-up tiling processes. Both types of models permit tunable disorder in topology, geometry, and other parameters. We also suggest that both types of models are amenable to be combined with topology optimization[100] when the latter can invoke homogenization. In this case, the network representation can match the mesoscopic structure of the constitutive elements, thereby facilitating scalable optimization by reducing the number of elements needed to accurately describe the material. Reducing the number of elements while maintaining a faithful material description is particularly helpful in nonconvex problems[101,102].

## Learning Approaches for Irregular Metamaterial Network Design

**From network models to deep neural networks**. A major technical challenge currently faced in designing irregular metamaterials with specified properties is—aside from specific simulation-driven optimization[35]—the lack of suitable methods for structure-property mapping, particularly when the system undergoes phase transitions[103], the network connectivity changes in time[104], and frictional forces are non-negligible[105]. By formulating irregular metamaterials as networks, they can be represented in a format conducive to training graph neural networks (GNNs) to infer this mapping, or to learn effective dynamical equations. The latter application translates to GNNs an idea successfully employed in convolutional neural networks[106,107] (CNNs) through graph convolutional networks[108]. Unlike CNNs, these models naturally allow for network data to serve as input, which includes the network topology, network geometry, and possibly other node and edge parameters (**Box 2**). GNNs are trainable on network examples that vary in size and remain invariant if the node labels are permuted[109]. Geometric symmetries can be accounted for by augmenting the training dataset with transformed configurations that have the same property labels or by embedding such symmetries in the neural network structure[110].

Deep neural networks have also been used to design desired nonlinear mechanical responses[111,112] and to infer structural properties from material imaging data[113,114]. The latter application is often carried out using a variational autoencoder (VAE) architecture. This unsupervised architecture is designed to uncover patterns in the image by encoding high-dimensional input data into a lower-dimensional latent space via an "encoder" neural network[115]. With the latent space in hand, a “property predictor” neural network is trained in a supervised manner using the latent variables as input and the material properties of interest as supervisory signals. These material properties can then be extracted from the latent space, and inverse design proceeds by finding points in the latent space corresponding to desirable properties. Selected points can be decoded into structures by using the latent variables as an input to the encoder run in reverse. The VAE architecture is generally operated on inputs of fixed size[113,114,116], although it is possible to combine them with GNNs to cope with variable-sized inputs[117].

Figure 7 highlights some recent materials applications of machine learning, including the prediction of: force chains[118] and particle rearrangements[119] in granular materials, where the network consists of nodes representing grains and edges representing grain contacts; heat transfer and stiffness in shell metamaterials, where truss networks dual to the shell surface are used both to construct surface and to serve as inputs to the neural network model[120]; and nonlinear stiffness in truss metamaterials, where the network consists of a point set connected by edges[116,121]. In all these examples, the machine learning model serves as an emulator for potentially computationally expensive finite-element calculations of homogenized properties. In cases where finite-element calculations are inexpensive, such as certain behaviors that are well-captured by the elasticity of the elements, direct simulation-based optimization can be preferred over neural network approaches[122,123].

GNNs have also proved useful to optimize force-directed layouts[124] and thus to assign geometry to abstract networks by embedding them in the physical space. Such layouts position nodes and edges to reduce overlaps between them. The process can be interpreted as identifying a hidden metric space, which is an underlying space wherein distances between nodes correlate with the probability that they are connected[125]. This offers a path to create physical networks with a balance of local and nonlocal interactions.

**Physical network learning.** Metamaterials can be recognized as physical networks, and such networks can be trained to learn a specific behavior by optimizing their parameters. Indeed, while traditional machine learning is a software implementation inspired by brain network learning, it is also the case that many other natural physical networks undergo learning, as illustrated in Fig. S2 for a protein network underlying embryo development[126]. A salient feature of “physical network learning” is that it is generally decentralized in the sense that the update rules are informed by local information, which stands in contrast to the global update rules used in software implementations of machine learning[127,128]. Crucially, when used for the design of metamaterial networks, such physical implementations of learning are expected to lead to the emergence of irregular edge and node parameters[129]. Indeed, the irregularity itself is the key to the solution of the inverse problem of having a desired response to a perturbation[130]. Such responses are often static, but learning dynamical responses is a current frontier in metamaterials research[131]. In addition to its significance to the design of metamaterials with exceptional properties, learning in irregular metamaterial networks can offer new platforms for hardware implementations of machine learning and artificial intelligence systems[132,133].

## Scalability and the Next Frontier of Manufacturing Approaches

**State-of-the-art and emerging manufacturing approaches.** The fabrication of metamaterials has traditionally relied on additive manufacturing (AM), which offers exceptional design freedom for creating architectures with a broad range of geometric complexity[134]. Commercial AM techniques have enabled the realization of large-scale periodic metamaterials in diverse material systems: polymers via stereolithography[135] or fusion-based techniques[136], ceramics via binder jetting or direct ink writing[137], and metals through laser powder bed fusion[138]. These methods have become standard in the field due to their reliability and scalability for macro-structured designs.

However, many of the most impactful metamaterial functionalities—particularly in optics, electronics, and mechanics—require architectures with nanoscale or sub-micron features. Optical and electronic metamaterials often exploit unit cell dimensions on the order of the wavelength of light to achieve tailored electromagnetic responses[139]. Similarly, mechanical metamaterials can harness nanoscale architectural features to enable ceramic constituents to approach their theoretical strength by suppressing flaw sensitivity and crack propagation[140]. These novel functionalities motivate the development of manufacturing strategies that produce macroscale samples from components of nanoscale resolution.

Two-photon polymerization direct laser writing (2pp-DLW) currently offers the highest spatial resolution available[141], and while historically confined to photosensitive polymers, its applicability has been dramatically expanded by recent advances in material conversion strategies. Techniques such as atomic layer deposition, pyrolysis, electroless deposition, electrodeposition, and ion infiltration followed by thermal processing have enabled the transformation of printed polymer scaffolds into ceramics[140], metals[142], composites[143], and even glasses[144]. These methods have produced metamaterials with record-breaking strength-to-weight ratios[3] and novel optical properties[144].

To overcome the inherent limitations of 2pp-DLW in terms of scalability, several promising developments have emerged. In particular, projection microscale lithography allows for faster fabrication of complex polymeric lattices over larger volumes[145], while new wide-area parallel DLW systems are pushing the boundaries of high-throughput nanoscale printing. As shown in Fig. 8, techniques that allow fabrication of structural components at the meter scale generally have resolutions of the order of several millimeters to centimeters (e.g., cold spray and wire-arc additive manufacturing), whereas processes that enable architectural features at the sub-micron scale (e.g., 2pp-DLW) are generally limited to sample sizes of a few millimeters (at least along two directions). Ultimately, all AM technologies inherently couple resolution and sample size, necessitating the development of new approaches based on self-assembly of the metamaterial from precursor components (blue dots in Fig. 8).

**Emerging self-assembly approaches to enhance scalability.** A fundamentally different emerging route to scalable metamaterial manufacturing leverages self-assembly followed by material conversion (red dots in Fig. 8). For instance, phase separation in bi-continuous interfacially jammed emulsion gels (bijels) and block copolymers have been used to template ceramic shell lattices with spinodal topologies, which are irregular yet structurally robust architectures[146]. Pushing these technologies to larger scales and improving the ability to tune the degree of topological disorder are frontier areas of research.

Cutting-edge self-assembly techniques can yield structures with sub-micron wall thicknesses and sizes extending to the centimeter or even meter scale. Despite being initially viewed as a drawback, the irregularity inherent to self-assembled structures has recently been recognized as advantageous, particularly in mechanical applications where it promotes failure delocalization and enhanced energy absorption under dynamic loading[147].

As novel disordered metamaterial architectures begin to emerge from network theory-driven design paradigms—whether truss-, shell-, or particle-based—they bring with them distinct challenges and opportunities for fabrication. One key challenge is to generate architectures with a highly tunable, application-specific type and degree of disorder. For structures in which control needs to be extended all the way down to nanometer scales, a particularly compelling research direction lies in tailoring the thermodynamics and kinetics of self-assembly processes. On the other hand, traditional AM-based techniques will remain uniquely capable for realizing disordered structures with exquisite control at macroscopic length scales. This opens opportunities for new hybrid approaches that combine self-assembly-based structuring at the smallest scales with AM-based fabrication at larger scales.

**Challenges and Opportunities**

Compared to regular metamaterials, irregular metamaterials have a larger parameter space. Since regular is a limiting case of irregular, this ensures that the optimum of a material property is at least as good as, and often superior to, that achievable by focusing exclusively on regular designs. The key challenge is performing property optimization while avoiding the search-space explosion typical of such problems. Regular metamaterials, being built from repetitive structures, require specification of only a single unit cell, whereas irregular metamaterials require specifying macroscopic portions—or even the entirety—of the system, which is then naturally represented as a network. Figure 9 shows recent examples of materials with unusual properties such as a tunable isotropic negative Poisson's ratio, enhanced resistance to damage, and both solid- and fluid-like behavior in the same material. These were designed by adopting a network perspective (Fig. 9c), by embracing irregularity (Fig. 9b), or both (Fig. 9a). As these examples show, leveraging a network representation is advantageous first and foremost because, all else being equal, structure (i.e., *the network*) determines function. Notwithstanding, the real opportunities lie in how networks can facilitate the design, optimization, and analysis of new metamaterials, which we categorize as follows:

***Forward problems****:* A common goal is to identify descriptors that characterize properties of given metamaterial realizations. The fact that additive descriptors, such as Minkowski functionals, do not explicitly account for correlations reflects a broader theme, namely that metamaterials exhibit nonadditive, emergent properties that arise from the interactions between building blocks. This underscores the necessity of holistic descriptors based on suitable network representations to characterize patterns arising from the differential structural arrangements. In aerospace applications, for example, nonadditivity can allow for the creation of stiffer and tougher metamaterials with the same weight by exploring the ensemble of possible irregular configurations. Identifying such network descriptors is an important direction in current research, as they can assist dimensionality reduction without discarding critical information about the interactions and correlations that ultimately govern the metamaterial's macroscopic behavior. The network perspective has been effective, for example, in characterizing material rigidity via an eigenvalue of a dynamical matrix representing the underlying network[148], as opposed to the more involved calculation based on Maxwell constraint count (and the associated determination of the null spaces of the compatibility and equilibrium matrices[23]).

***Inverse-forward-hybrid problems****:* Network modeling is especially well-suited to predicting material behavior or properties using a simplified representation rather than the complete physical description. For example, algorithms have been developed to identify, in an arbitrary network, the "communities" of densely connected nodes that are sparsely connected with each other—a network technique previously applied to characterize material structure[149]. An approach to identify community structure in a network is by partitioning it based on the so-called betweenness centrality of the edges, and then sequentially removing the edges with the largest centrality[58]. The betweenness centrality of an edge is the fraction of shortest paths between nodes in the network that pass through that edge and is thus a measure of the edge's importance in connecting different parts of the network. A similar calculation can successfully predict the sequence of bond breaking (and thus the most critical network components) when a metamaterial network is loaded, as previously shown for disordered lattices[150]. In addition to its computational advantages, this approach illuminates the relationship between a material's internal structural organization and its failure pathway—insights that are not immediate from direct finite-element simulations.

Analogous network-based surrogate prediction approaches have achieved remarkable success in biomedical[151] and sociotechnological[152] systems. A promising frontier in irregular metamaterials research is thus to develop such approaches to enable accurate predictions without recourse to exhaustive direct (e.g., finite-element or molecular dynamics) simulations, which can be impractical, as is often the case when the deformations of a material are large, the structural topology is dynamic, and friction is non-negligible.

***Inverse problems:*** The overarching goal in exploring a larger space of metamaterial structures is to design materials with desired properties and responses, including unprecedented ones. As argued above, a promising approach to address such design problems is GNNs, and network representations are uniquely suited to serve as a bridge to the incorporation of data and/or simulations into such machine learning models, which yield a "network structure-material property" mapping, which can be inverted to obtain network structures that achieve the desired properties. The fact that the output of this process is a network is advantageous because the network abstraction is sufficiently flexible to incorporate a wide range of material structures of varying size and even composition, meaning that GNNs can be trained on diverse samples. These models can be particularly well-suited to leverage network structure in the optimization of hard-to-calculate properties, such as fracture toughness. This is because GNN training involves recursive message passing on the underlying network, which means that the learned message functions encode how local interactions are aggregated and propagated across network neighborhoods, allowing local structural behavior to influence properties of the entire material[153,154].

The advantages of irregular over regular metamaterials, and of representing such materials as networks, can be further appreciated by considering the effectiveness of artificial neural networks in machine learning. In both cases, irregularity expands the parameter space, enabling the representation of a wider range of possible behaviors. To optimize specific material properties, the parameters are irregular but not random. This is where a crucial advantage of the network representation arises, in addition to those already noted above: it transforms irregular data into structured data that preserve correlations and suppress irrelevant variations, thereby facilitating interpretability, dimensionality reduction, and generalization. This power can lead to advances on multiple fronts:

*Multifunctionality*: The enlarged design space of irregular metamaterial networks, and resulting increased number of control inputs, opens the possibility of systematically encoding multiple disparate or competing[2] properties within the same material.

*Programmability:* By the same principle, such networks enable the programming[155] and reprogramming[156] of distinct responses to specific external stimuli, creating new opportunities for dynamically tunable, modulable, reconfigurable, and adaptive metamaterials.

*Performance with robustness*: A major goal is to realize desirable functional properties in combination with robustness to aging and failure[157,158]. The structure-property mapping of irregular metamaterial networks supports the simultaneous pursuit of both functional performance and robustness.

*Multistability and inelastic behaviors*: Irregular networks can be designed to exhibit multiple stable states with target properties, which in turn can lead to unique nonlinear inelastic responses—such as negative compressibility transitions[103]—that are absent in monostable systems.

*Extreme and unprecedented properties*: Ultimately, an enlarged property space produces a structure–property mapping that is more complex, higher-dimensional, and harder to explore, but also richer in potential functionalities[159].

It is thus clear that irregular network-based designs are advantageous not only for enhancing specific functions, such as energy dissipation, but also for improving robustness against localized failures. Broader exploitation of irregular metamaterial networks holds the potential to advance emerging areas, including topological, transduction, neuromorphic, active, biocompatible, microfluidic, space-time, and quantum metamaterials, among others. At the same time, challenges such as managing the combinatorial explosion of parameters, training machine learning models with limited data, and developing theory-guided approaches all highlight the value of a network-theoretic formulation.

## Final Remarks

The structure of a metamaterial network comprises topology, geometry, and the nature of the network components, which are generally interdependent. For example, specific topological features emerge as a consequence of the spatially embedded nature of these networks, as illustrated in our geometric extension of the Erdős–Rényi model and observed in empirical physical networks[160,161]. This property of physical networks transcends metamaterials and architected materials in general[162], as exemplified by the recent suggestion that the network organization of the human brain itself may be strongly influenced by geometric constraints[163,164]. Previous research has shown the significance of network concepts in designing new metamaterial behavior[103], in modeling coordinated motion of connected material parts[21] and hierarchical structural organization[92], and in characterizing the impact of geometric constraints[20,89]. The field is now primed to develop a comprehensive network theory for systematically studying the space of possible *irregular* metamaterials with unprecedented properties.

## Acknowledgements

The authors thank Chelsea Fox for providing images of metamaterial samples, and Doruk E. Gokmen, Kate M. Ainger, Ethan Stanifer, and Yi Zhao for insightful discussions. This work was supported by MURI ARO grant No. W911NF-22-20109.

## Author Contributions

TPW, CD, HMJ, CAS, VV, LV, and AEM conceived the content of the article.

TPW and AEM led the writing with contributions from all authors.
All authors reviewed and approved the final version.

**Competing Interests**
The authors declare that no competing interests exist.

## List of Figures

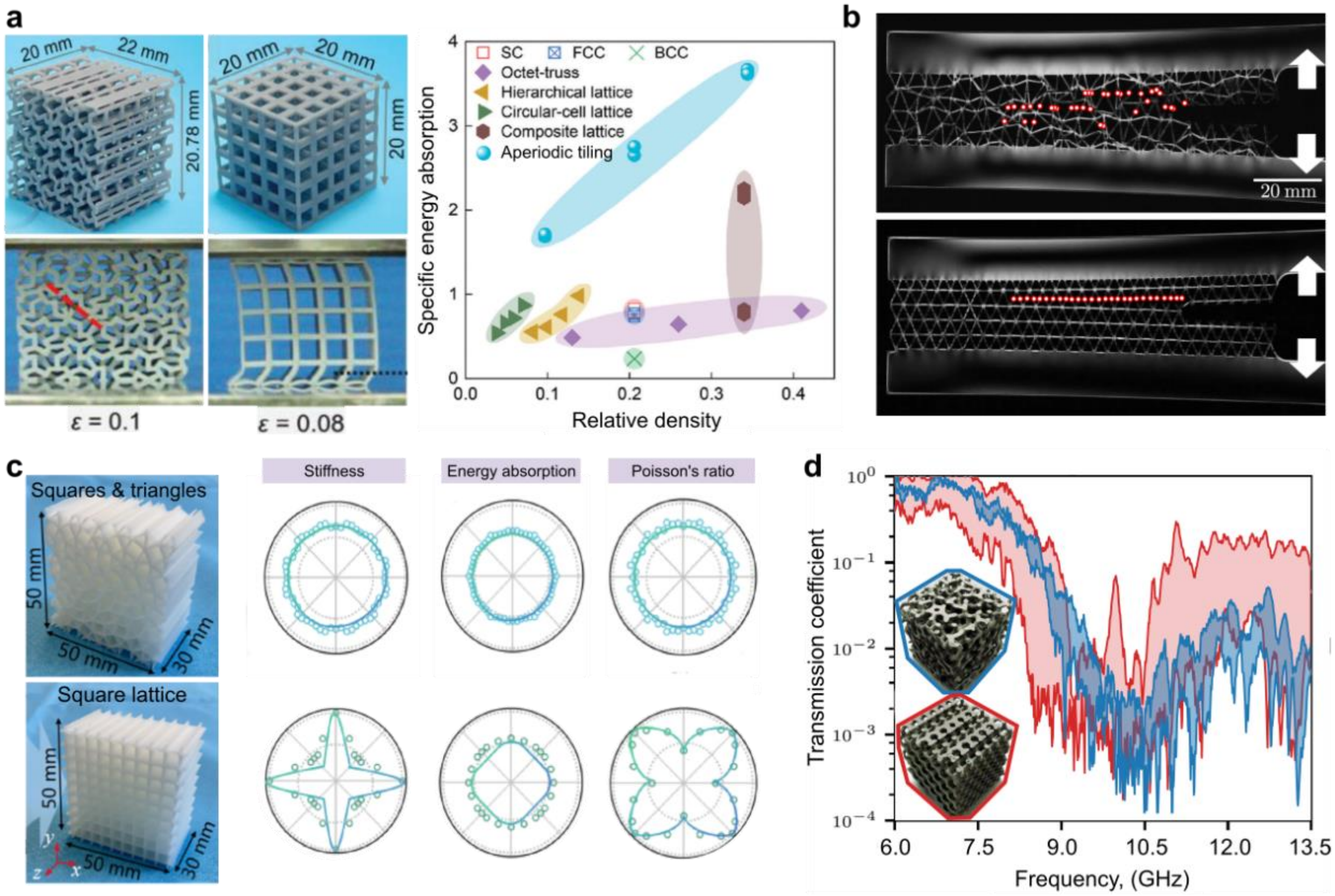

**Fig. 1: Examples of exceptional behavior achieved by irregular metamaterials. a,** Samples of aperiodic (top left) and simple cubic (top right) structures, their behavior at approximately 10% strain (bottom left and right, respectively), and their specific energy absorption as a function of density (far right)[165]. **b,** Fracture propagation in an irregular truss structure (top) versus its regular counterpart (bottom), when the right side is pried apart (white arrows). The irregular structure deflects the crack tip, leading to a doubling of the fracture toughness[157]. **c,** Isotropy of the stiffness, specific energy absorption, and Poisson's ratio for an aperiodic lattice of squares and triangles (top row) compared with the anisotropy of these characteristics for the square lattice (bottom row)[166]. **d,** Approximate isotropic light transmission behavior of an irregular hyperuniform structure (blue) compared to the strongly anisotropic behavior of the diamond lattice structure (red), where the colored area spans the min-max range of transmission coefficients at a given frequency across the three principal directions[167]. The changes in the transmission coefficients reflect changes in the photonic bandgap defined by the given structure. The images are adapted with permission from the cited references.

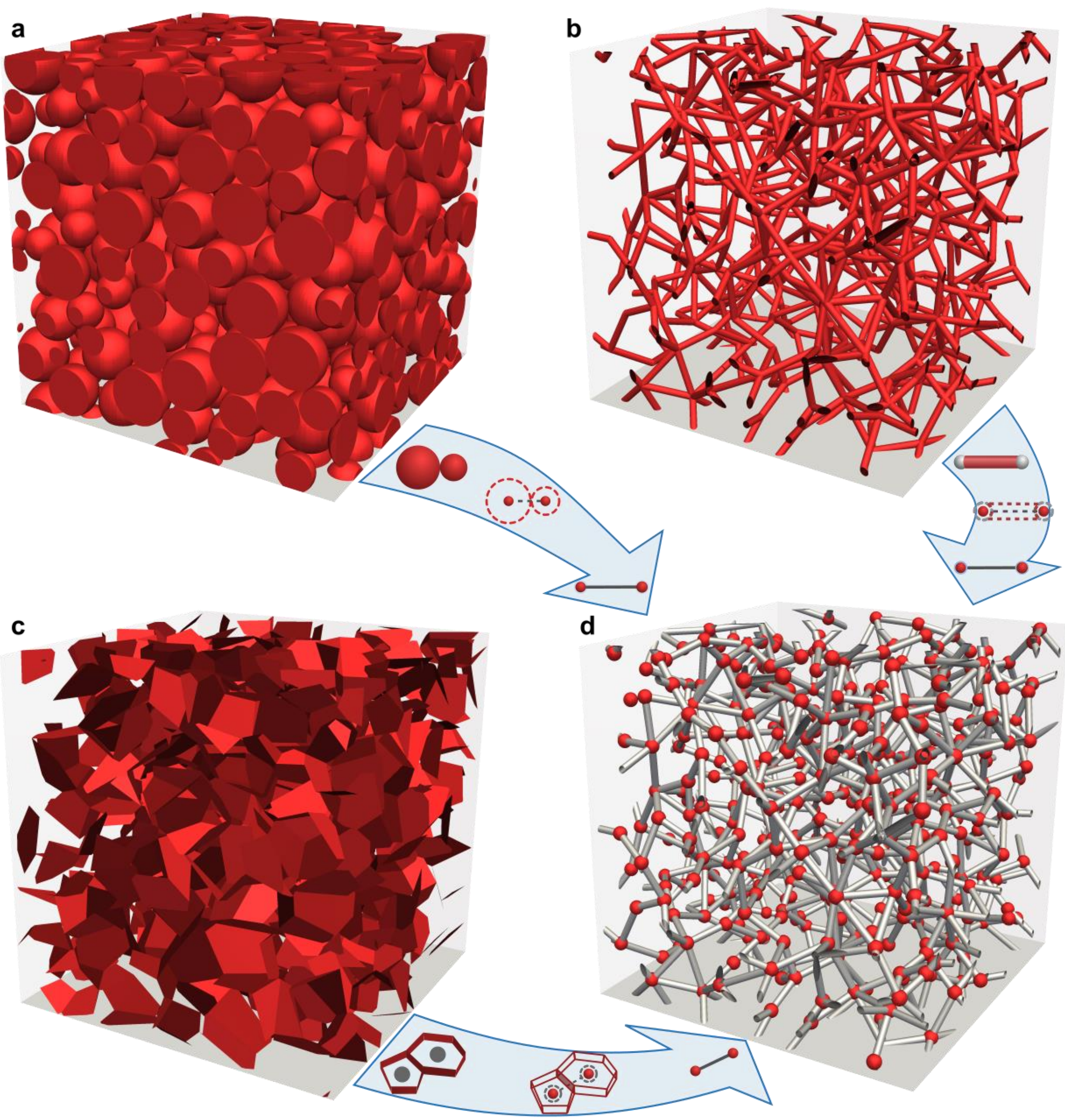

**Fig. 2: Schematics of metamaterials that can exist in periodic and irregular form. a-c**, Irregular granular, truss, and shell structures, respectively, corresponding to interactions of increasing dimension and complexity. **d**, Underlying network topology, which is the same in (a-c) for comparison. The arrows indicate the relationship between the network topology and material elements in each case (additional details on the network construction are provided in the Supplementary Information). The image is inspired by Ref. 168.

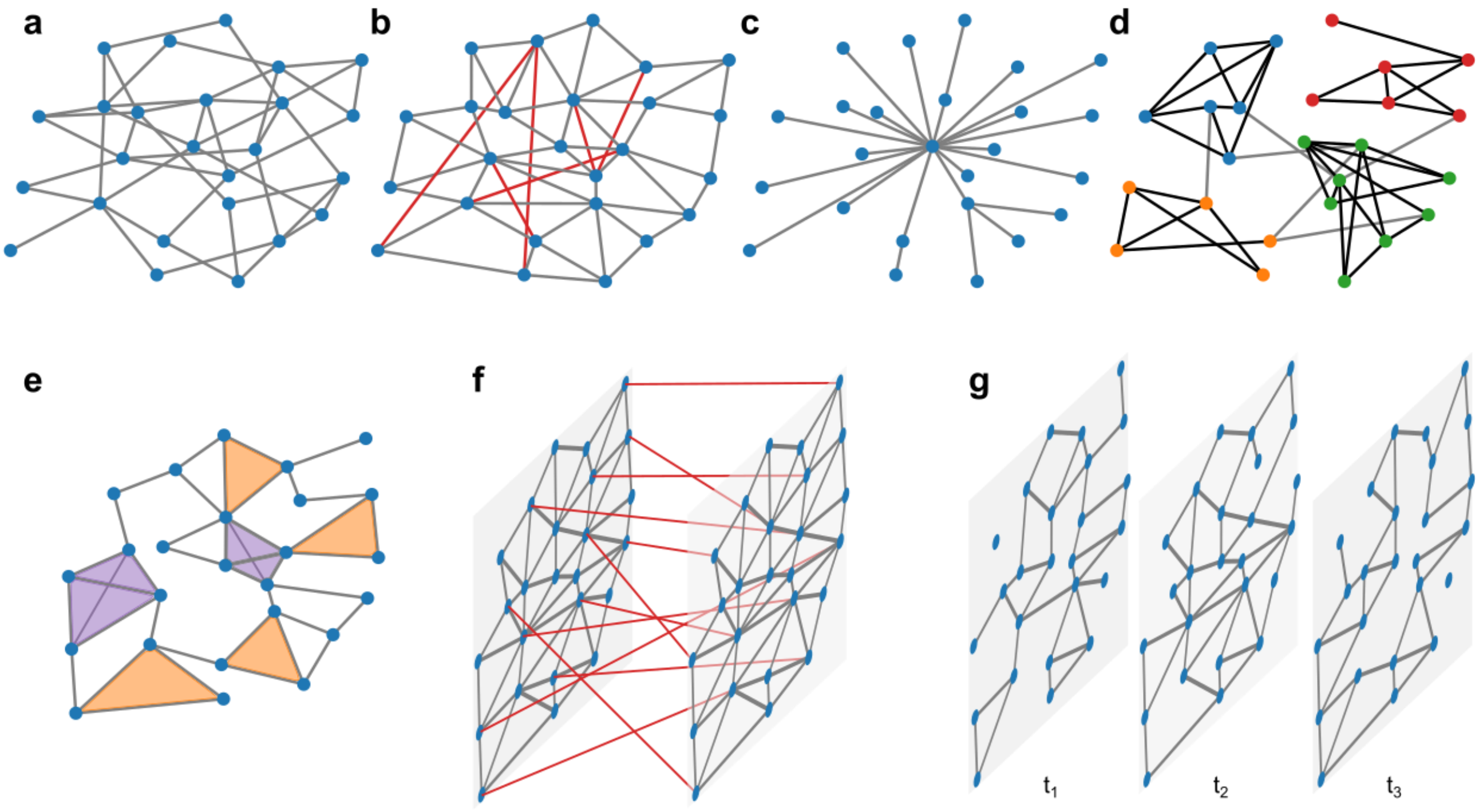

**Fig. 3: Irregular networks generated by existing models.** **a**, Erdős–Rényi network, where node pairs are connected by an edge with a given probability $p$. **b**, Small-world network, generated through the rewiring of a small fraction of edges in a triangulated lattice. **c**, Scale-free network, generated by randomly connecting nodes with a pre-assigned power-law distribution of the number of connections. **d**, Network with communities of densely connected nodes generated using the stochastic block model with four planted clusters. **e-g**, Generalized networks with higher-order (e), multilayer (f), and temporal interactions (g).

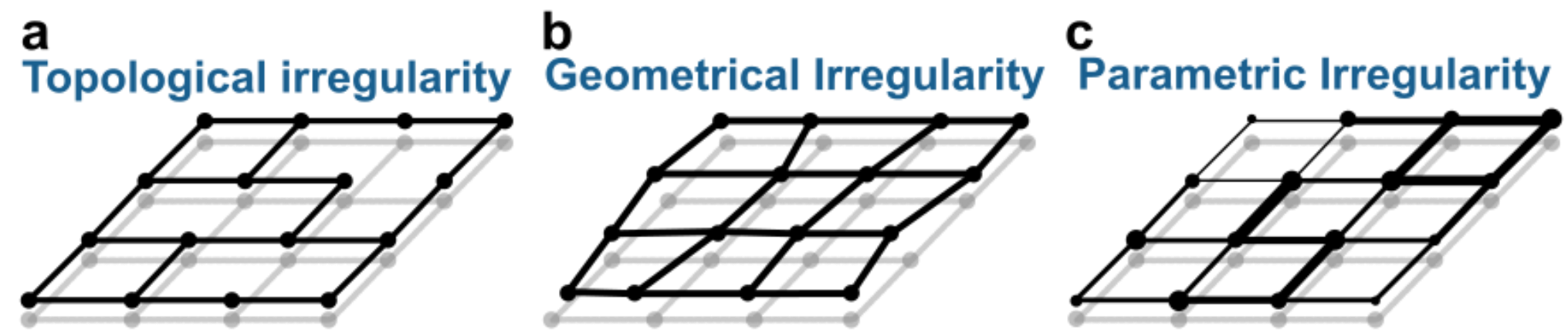

**Fig. 4: Types of irregularity. a-c,** Illustration of topological (a), geometrical (b), and parametric irregularity (c).

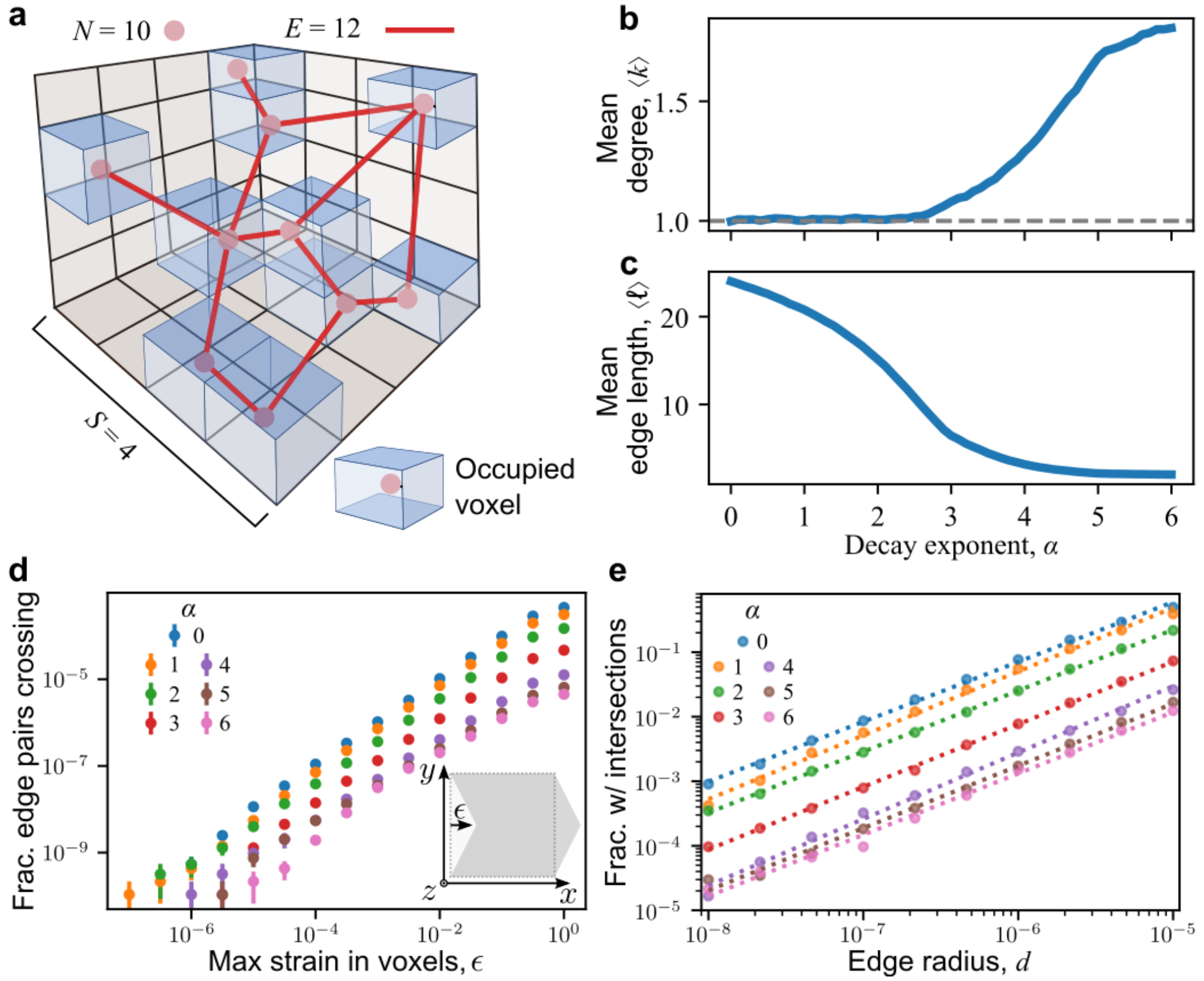

**Fig. 5: Geometric extension of the Erdős–Rényi model. a**, Example network for $N = 10$ nodes, $E = 12$ edges, and $S = 4$ voxels per side of the principal volume. **b**, Mean degree at the percolation transition as a function of the decay exponent $\alpha$ for edge radius $d = 0$, which has qualitatively similar behavior to the model in Ref. 88. The dashed line indicates the degree in the (topological) Erdős–Rényi model. **c**, Mean length of the edges for the networks in (b). **d**, Crossings in a strained material for $d = 0$ as a function of the maximum strain (in units of voxel side lengths) for the profile shown in the inset. **e**, Fraction of edges violating the volume exclusion constraint as a function of $d$ in the absence of loading. In (b-e), $N = 6250$ and $S = 50$; in (d), error bars larger than the symbol denote the standard error of the mean.

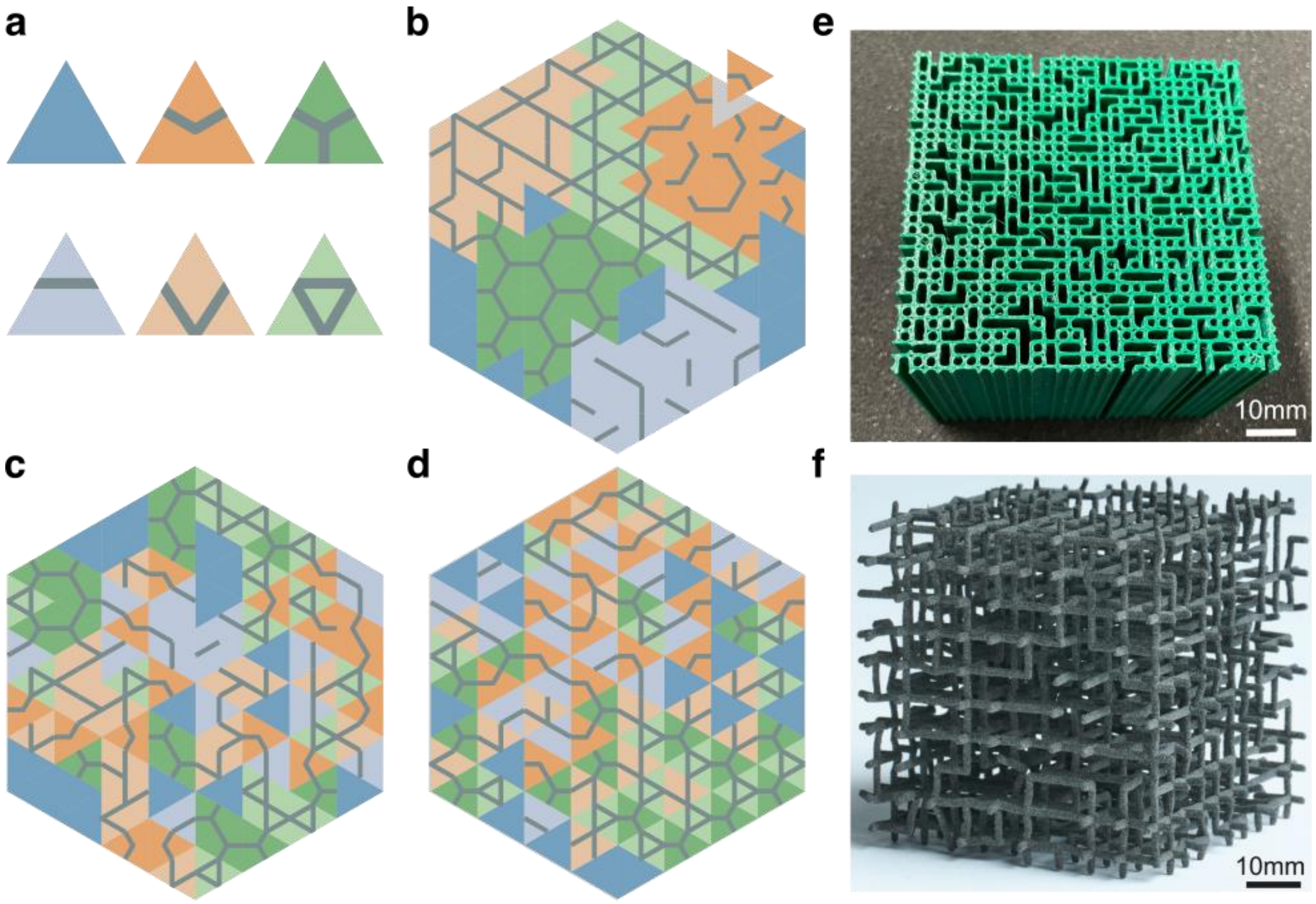

**Fig. 6: Network model and metamaterials based on virtual growth. a,** Example building blocks. **b-d,** Model realizations for the same proportions of the building blocks and material connections with different short-range correlations: (b) positively correlated, (c) uncorrelated, and (d) negatively correlated. **e-f**, 3D-printed realizations for (e) 2- and (f) 3-dimensional metamaterials with different building blocks (image in (e) from Chelsea Fox, image in (f) reproduced from Ref. [96] with permission).

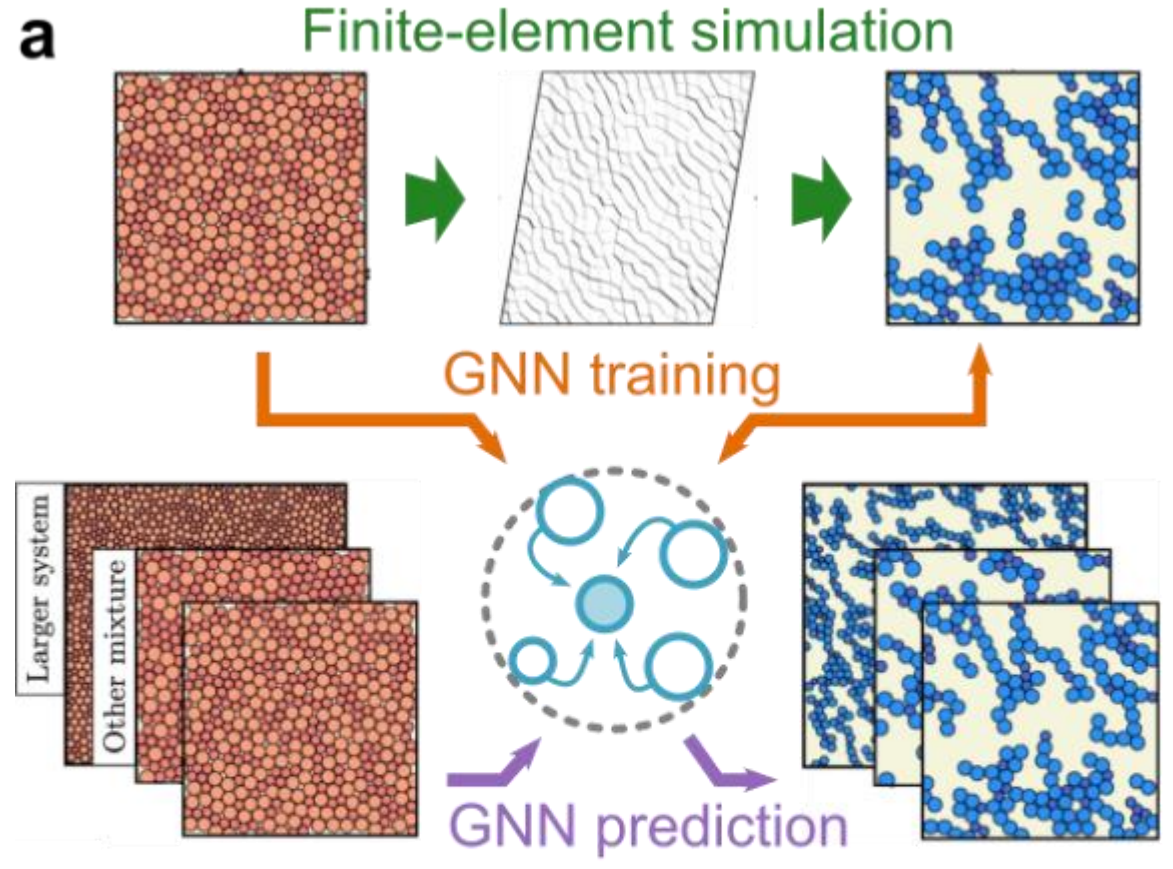

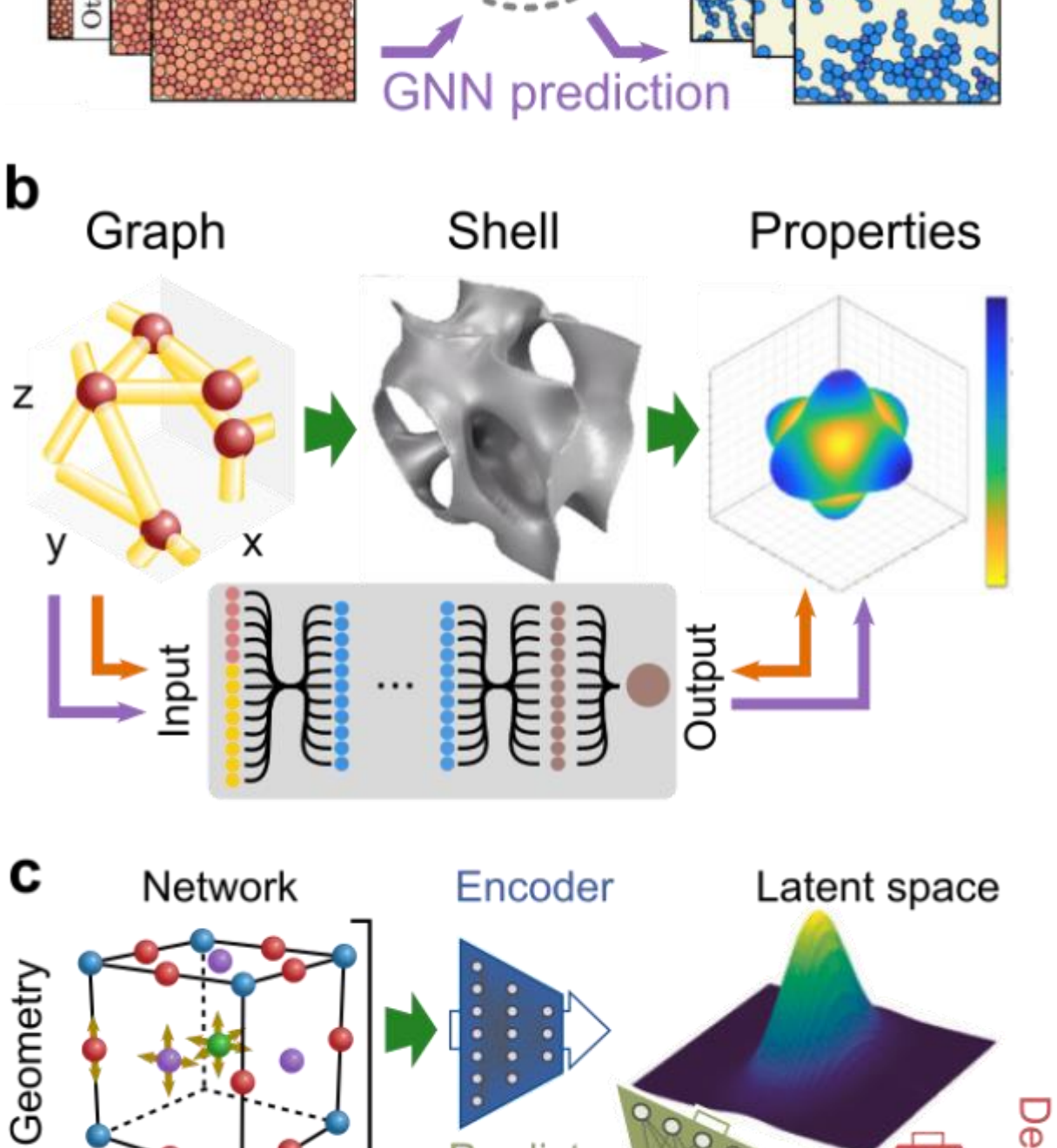

**Fig. 7**: **Neural networks for metamaterial design. a**, Prediction of force chains in jammed packings of grains using a GNN trained on finite-element simulations[118]. **b**, Graph representation of shell-metamaterials used to train GNN-based structure–property models[120]. **c**, Variational autoencoder framework to design truss-based metamaterials with tailored nonlinear behaviors trained on network topology and geometry[116]. The images are adapted from the cited references with permission.

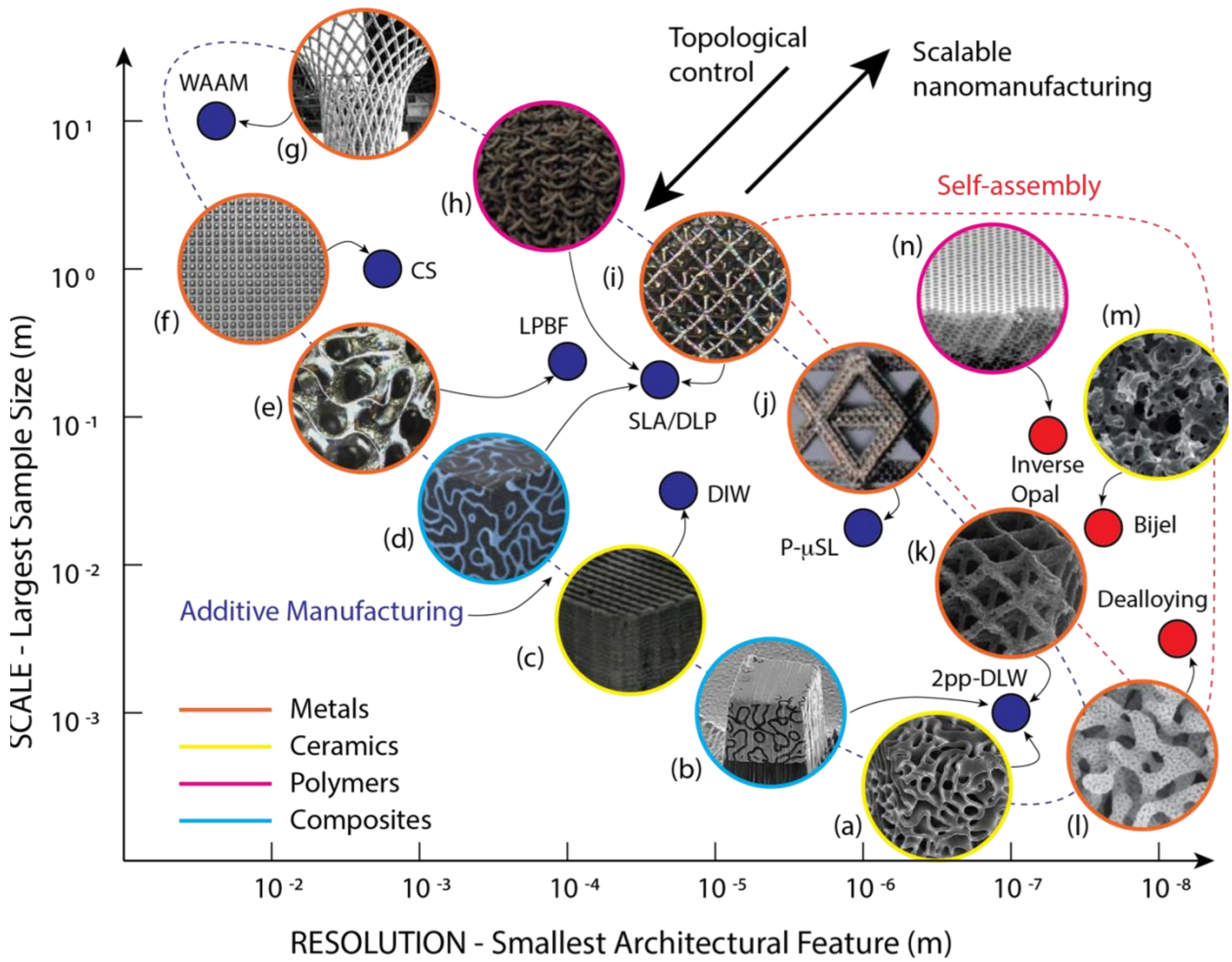

**Fig. 8: Inverse relationship between scale and resolution across AM technologies. a-n,** AM technologies (blue dots) and self-assembly processes (red dots) can fabricate periodic and disordered metamaterials of various constituent classes, including metals, ceramics, polymers, and composites: carbon spinodal nanolattice[140] (a), carbon/nickel interpenetrating phase nanocomposite[143] (b); ceramic woodpile structure[137] (c); polymeric interpenetrating phase composite[147] (d); medium entropy alloy gyroid metamaterial[169] (e); 2D array of aluminum heat sink fins[170] (f); aluminum architected structure[171] (g); polymeric polycatenated architected material[172] (h); metallic microlattice[173] (i); multiscale metallic metamaterial[174] (j); metallic nanolattice[142] (k); hierarchical gold spinodal nano-structure[175] (l); graphene spinodal shell metamaterial[146] (m); and ceramic inverse opal structure[176] (n). The state-of-the-art methods to produce these materials include two-photon polymerization direct laser writing (2pp-DLW), projection micro-stereolithography (P-μSL), direct ink writing (DIW), stereolithography (SLA), digital light processing (DLP), laser powder bed fusion (LPBF), cold spray (CS), and wire arc additive manufacturing (WAAM). All material images are reproduced from their respective references with permission.

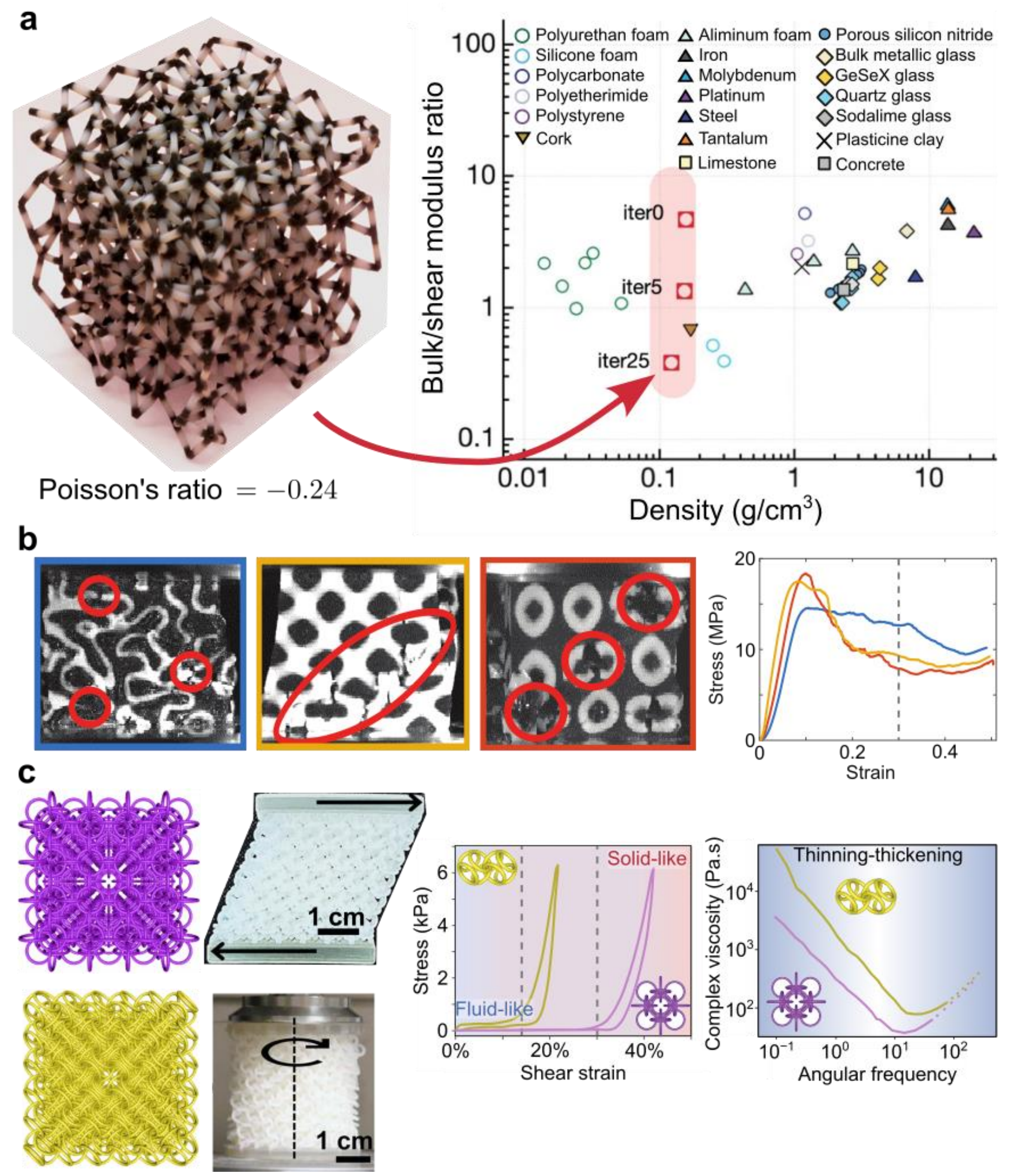

**Fig. 9: Irregular metamaterials with underlying networks possessing unusual properties. a,** Truss metamaterial exhibiting a 3D isotropic negative Poisson's ratio that is tuned by optimizing the locations of the truss intersections[177]. **b**, Two-phase composite spinodal shell metamaterial with enhanced energy absorption and damage resistance[147]. **c,** Polycatenated metamaterial exhibiting both solid-like and fluid-like behavior under shear, as well as shear-thinning and -thickening under torsion[172]. The irregularity in (a) and (b) is designed, whereas in (c) it is induced by gravitational relaxation on an otherwise regular underlying network arrangement. The images are adapted from the cited references with permission.

**Box 1 | Emerging models of physical networks**

We comment on two growing classes of physical network models of relevance to metamaterials. The first class extends the traditional notion of networks by embedding nodes in 3D space and drawing edges between them. In one model, edges of defined thickness are sequentially added until no more edges can satisfy volume exclusion (see figure, panel **a**)[19,20]. The latter model uses the concept of a "metagraph" to satisfy volume exclusion, thereby mapping the problem of satisfying edge volume exclusion to the identification of independent sets on graphs. Another model focuses on how nodes and edges embedded in space form knots (panel **b**), which can be related to the behavior of complex liquids[89,178,179], solids[180], and gels[181]. Similar work on physical entanglement has been recently applied to networks of fibers[47,182,183]. In a third model, the nodes themselves are chain subnetworks, which are grown on a defined grid until they run into another node, forming an edge (panel **c**)[92]. These and related models can describe long-range interactions, producing structures that satisfy volume exclusion while edges and/or nodes are physically extended.

The second class uses Maxwell frames—rigid rods of fixed length (edges) that connect at hinge joints (nodes)—to prescribe the motion of target nodes in response to the actuation of distant source nodes. An early model in this class was used to design 2D and 3D frames that exhibit allostery, where the target nodes move apart in response to the source nodes being pinched together (panel **d**)[22]. This model illustrates how, even in locally coupled networks, architected disorder can be used to program nonlocal responses in metamaterials beyond those mediated by wave propagation[184]. Further work explored determining placement of source nodes to achieve a prescribed deformation pattern for a given network (panel **e**)[21] and composing networks consisting of a series of motifs that achieve prescribed folding patterns (panel **f**)[185]. These models can successfully program specific mechanical behaviors by tuning the short-range interactions (existence and length of the edges between nearby nodes). All images are adapted from their respective references with permission.

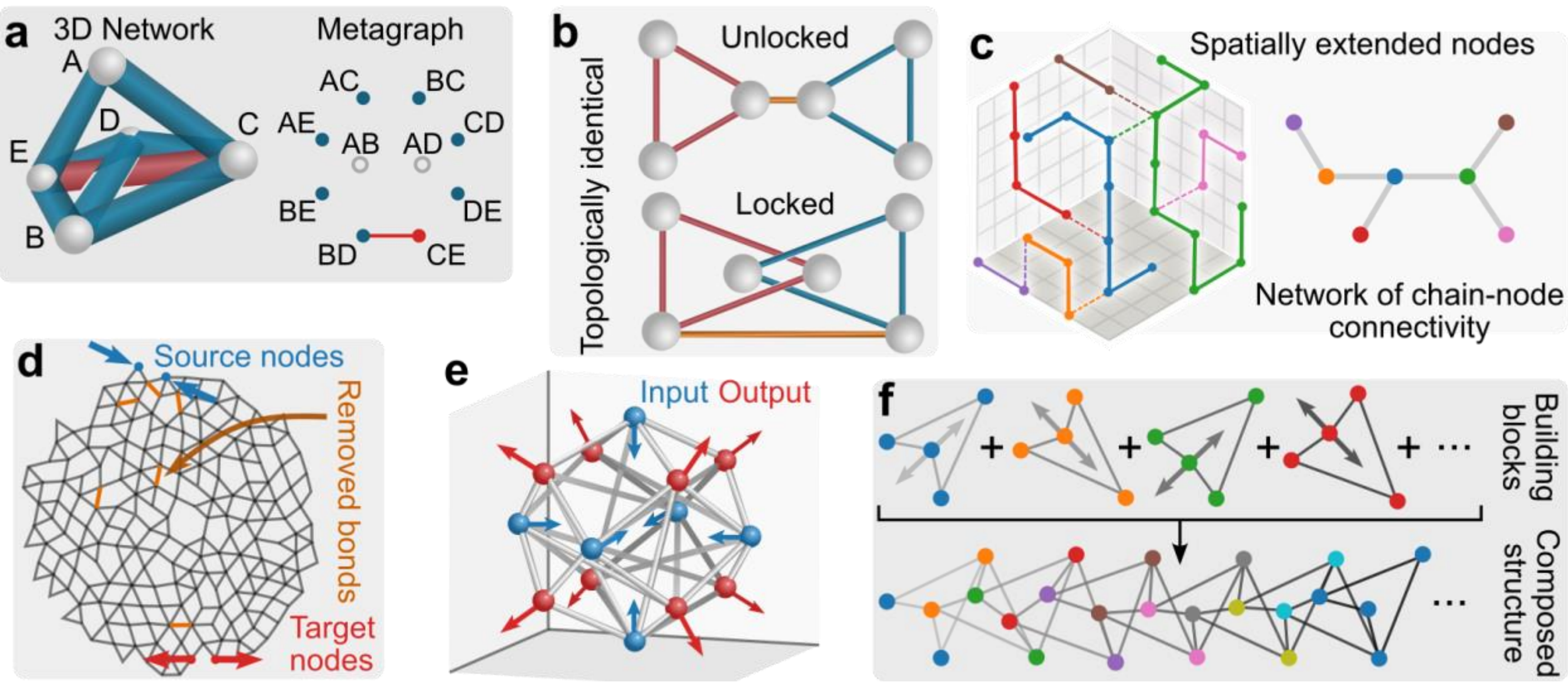

**Box 2: Graph neural networks for metamaterial design**

The figure describes the information processing of CNNs and compares it with that of GNNs. As illustrated in panels **a** and **b,** a CNN can be used to construct a latent representation (blue vector) of an image by taking a weighted average of pixel values using a trained filter. This can be equivalently formulated as a special type of message passing operation on a regular lattice graph (panel **c**), where each node carries an RGB feature vector. The underlying operation can be interpreted as a coarse-graining that extracts relevant local descriptors[153,154]. However, while CNNs are generally designed for the analysis of continuum systems[106,107]—in which information between nodes is shared locally in space (e.g., image classification)[186] or time (e.g., speech processing)[187]—the message passing formulation generalizes to discrete systems, including irregular networks as seen in GNNs[188].

Message passing consists of applying a *message function,* an *aggregation operator*, and an *update function* throughout all edges of the network. The message function takes attributes attached to an edge along with the attributes associated with the nodes it connects, and it generates an output message. The aggregation function combines these messages from all edges incident on a node and has two important properties: it accepts an arbitrary number of arguments, and it is invariant with respect to permutations of its arguments, which are together referred to as equivariance. The update function takes the aggregated message output at a node and uses it to update the node representation. After a specified number of steps, the node representations are fed into a *readout layer,* which maps the node representations to the desired output (e.g., a material property like stiffness or a node property like failure).

GNNs focus on training the message function, update function, and readout layer[189]. As illustrated through the example of polydisperse granular media in panels **d** and **e**, irregular networks are readily obtainable from material data. The nodes in such a network can have multiple degrees of freedom, including particle positions and other attributes (e.g., particle radii). Message passing is implemented on the whole network structure (panel **f**), allowing the node representation to evolve according to the state of its neighbors in the graph. The equivariance of the aggregation function and readout layer enables the same GNN to be trained on networks of different sizes, as is common in irregular network metamaterials. Figure courtesy of Efe Gokmen.

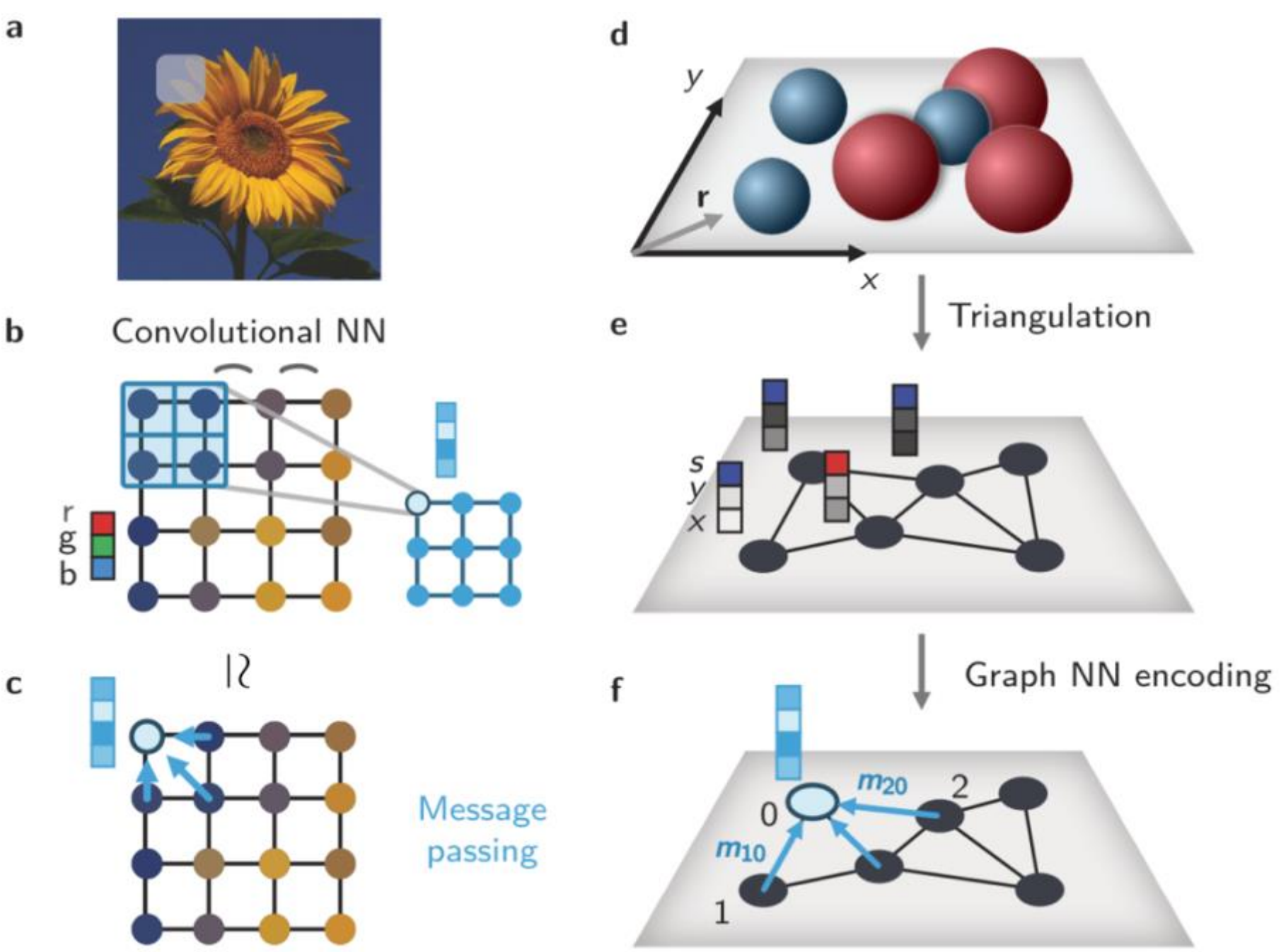

Supplementary Information

# Irregular Metamaterial Networks

Thomas P. Wytock, Chiara Daraio, Heinrich M. Jaeger, Christopher A. Schuh, Lorenzo Valdevit, Vincenzo Vitelli & Adilson E. Motter

## Table of Contents

### S1. Details of the 3D Erdős–Rényi model

In this section, we detail the approach taken to generalize the topological Erdős–Rényi model to the geometrically endowed version that accounts for the cost of adding each edge. We recall that the (topological) Erdős–Rényi model provides a scheme for generating random networks with $N$ nodes and an expected number of edges $E = p[N(N-1)]/2$, where each pair of nodes is connected independently with the same probability $p$. This model is modified by embedding the nodes in space and allowing the wiring probability $p$ to decay with the distance between the given nodes. Specifically, we assume that $p \propto r_{ij}^{-\alpha}$ asymptotically, where $r_{ij}$ is the *Manhattan distance* (a.k.a. city block or taxicab distance) between the voxels containing nodes $i$ and $j$, and $\alpha$ is the decay exponent. Fig. 5a illustrates the cubic principal volume of $S^3$ cubic voxels in which the nodes are embedded, with periodic boundary conditions imposed to eliminate boundary effects. We apply the minimum image convention[1] when computing the distance between nodes $i$ and $j$: node $i$ (in the principal volume) is connected to the nearest copy of node $j$, even if it resides outside the principal volume.

We distribute the $N$ nodes uniformly randomly into the $S^3$ voxels, with no more than one node per voxel. Each node has the same radius $R$, and it is placed uniformly randomly within the voxel (i.e., no part of the node extends outside the voxel to which it is assigned). These choices are made both to prevent nodes from overlapping and to maintain computational tractability by enabling intersections between objects to be checked on a per-voxel basis. Then, for each pair of nodes $(i, j)$, an edge is placed with probability $p = \min\{Ar_{ij}^{-\alpha}, 1\}$, where $A$ is a normalization constant and the minimum enforces the condition $p \leq 1$. For $\alpha \to 0$, we recover $p = A$, which is the same as that observed in the original Erdős–Rényi model.

We calculate the expected percolation transition by replacing $p$ with the ensemble-averaged connection probability $\langle p \rangle = A \sum_{r=1}^{3S/2} n(r)/r^{\alpha}$, where $n(r)$ is the number of sites a Manhattan distance $r$ away from the site occupied by a given node. Aside from replacing $p$ by $\langle p \rangle$, the derivations remain the same as in the original Erdős–Rényi model for the size of the largest connected component at the percolation transition when $A \leq 1$ and for the probability of forming a single connected component for any $A$. These arguments can be found elsewhere[2,3], so we do not reproduce them here in full. We merely note that (i) the percolation threshold is derived by assuming that the network is tree-like and (ii) the threshold to becoming a single connected component is derived assuming that this transition occurs by joining a component of $N-1$ nodes with an isolated node. This means that $\langle p \rangle = 1/N$ at the percolation transition and $\langle p \rangle = \ln N/N$ at the transition to a single connected component. When $A > 1$, nodes within a finite neighborhood $z = \lfloor A^{1/\alpha} \rfloor$ are connected with probability 1, where $\lfloor \cdot \rfloor$ denotes the greatest integer or floor function. This causes the network to no longer be tree-like, leading to excess edges and a corresponding increase in $\langle p \rangle$ (and thus the average degree $\langle k \rangle$) at the threshold. On the other hand, the value of $A$ does *not* impact the validity of condition (ii). As $\alpha$ increases, we observe a strong suppression of edges with $r > z$, which is consistent with convergence toward $z$-compact neighborhood site percolation on cubic lattices[4] as $\alpha \to \infty$.

In Fig. 5b,c, for each value of $\alpha$, the constant $A$ is chosen such that the network is at the percolation threshold. For values of $\alpha$ where $A \leq 1$, the giant component has $N^{\frac{2}{3}}$ nodes as in the Erdős–Rényi model. For large values of $\alpha$ (and thus for $A \gg 1$), the spanning cluster is expected to scale with a larger exponent, since the scaling is approximately $N^{\frac{5}{6}}$ for 3D simple cubic site percolation[5]. Therefore, we can compute a lower bound for $\langle k \rangle$ at percolation by choosing $A$ such that the size of the largest connected component is $N^{\frac{2}{3}}$. In Fig. 5b, the average degree is $\langle k \rangle = 1$ as $\alpha \to 0$, which implies $A = 1/N$; as $\alpha$ increases, a giant component is still formed with $\langle k \rangle = 1$ until $A > 1$ around $\alpha \approx 2$. Above this value of $\alpha$, the average degree $\langle k \rangle$ increases from 1. This trend is qualitatively similar to that observed in a 2D spatial network model[6]. Fig. 5c shows that the average edge length $\langle \ell \rangle$ decreases as a function of $\alpha$, as expected, capturing the behavior of a geometric aspect of the network. In this figure, $\ell_{ij}$ is the *Euclidean* rather than Manhattan distance between nodes $i$ and $j$. This choice is made because $\ell_{ij}$ is proportional to the amount of material needed to manufacture a strut. The trend is much more sharply peaked compared to a previous 2D spatial network model[6].

In Fig. 5d,e, the value of $A$ is chosen such that the network consists of a single connected component. Our analysis shows that the number of edges to form a single connected component remains at $N \ln N$, which we verified numerically for $\alpha$ in the range from 0 to 6. Figure 5d shows results for edge crossings in the limit $d \to 0$ ($d$ is the edge radius) when the mid-plane of the cube is strained as illustrated in the cartoon. In this limit, *affine* strains do not lead to edge crossings, which is why a non-affine strain is used.

We determine edge crossings by computing the signed distance between edges in the unstrained and maximally strained configurations and observing whether the distance changes sign. For edge pairs for which the sign changes, we write conditions to ensure that the point of intersection between the two lines lies within the segments occupied by the edges. To compute the signed distance, we recall that the edge between nodes located at points $\boldsymbol{p}_i$ and $\boldsymbol{p}_j$ in the unstrained configuration is described by the parametric equation $\boldsymbol{a}_{ij} + s_{ij}\boldsymbol{b}_{ij}$ for $s_{ij} \in [-1/2, 1/2]$, where $\boldsymbol{a}_{ij} = (\boldsymbol{p}_i + \boldsymbol{p}_j)/2$ and $\boldsymbol{b}_{ij} = (\boldsymbol{p}_j - \boldsymbol{p}_i)$. The signed distance between the entire lines defined by edges $(i,j)$ and $(k,l)$ is given by $D_{ij,kl} = (\boldsymbol{b}_{ij} \times \boldsymbol{b}_{kl}) \cdot (\boldsymbol{a}_{k\ell} - \boldsymbol{a}_{ij})$. Furthermore, the value of $s_{ij}$ at which the minimum distance $D_{ij,kl}$ is achieved is

$$s_{ij} = \frac{\boldsymbol{b}_{kl} \times (\boldsymbol{b}_{ij} \times \boldsymbol{b}_{kl}) \cdot (\boldsymbol{a}_{kl} - \boldsymbol{a}_{ij})}{\left(\boldsymbol{b}_{kl} \times (\boldsymbol{b}_{ij} \times \boldsymbol{b}_{kl})\right) \cdot \boldsymbol{b}_{ij}} \,,$$

and $s_{kl}$ is similarly obtained by exchanging $\boldsymbol{b}_{kl} \leftrightarrow \boldsymbol{b}_{ij}$ and $\boldsymbol{a}_{kl} \leftrightarrow \boldsymbol{a}_{ij}$. The coordinates, distances, and crossing points are functions of the strain $\sigma$. The coordinates transform as $\boldsymbol{p}_i(\sigma) = \left(p_{x,i}(\sigma), p_{y,i}(\sigma), p_{z,i}(\sigma)\right)$, where $\boldsymbol{p}_i(0) = \boldsymbol{p}_i$ and $\boldsymbol{p}_j(0) = \boldsymbol{p}_j$, and

$$p_{x,i}(\sigma) = p_{x,i}(0) + \sigma\left(1 - 2\left|\frac{p_{y,i}(0)}{S} - \frac{1}{2}\right|\right), \qquad p_{y,i}(\sigma) = p_{y,i}(0), \qquad p_{z,i}(\sigma) = p_{z,i}(0).$$

The vectors $\boldsymbol{a}_{ij}(\sigma)$ and $\boldsymbol{b}_{ij}(\sigma)$ are determined by substituting in $\boldsymbol{p}_i(\sigma)$ and $\boldsymbol{p}_j(\sigma)$ for $\boldsymbol{p}_i$ and $\boldsymbol{p}_j$, respectively, and the distances $D_{ij,kl}(\sigma)$ and crossing points $s_{ij}(\sigma)$ are then computed as described above.

An edge crossing occurs at $\sigma = \sigma^*$ when $D_{ij,kl}(\sigma^*) = 0$ and both $s_{ij}(\sigma^*)$ and $s_{kl}(\sigma^*)$ are in the interval $[-1/2, 1/2]$.

When computing overlaps between two edges and between one node and one edge, we assign the radius of the node and the radius of the edges to be the same value $d$, and we determine the voxels to which each node and edge belong. Then, for each voxel with more than one object (where at most one of them can be a node), we determine whether each pair of objects overlaps. Figure 5d,e shows, respectively, that the fraction of edges crossing as a function of strain and the fraction of objects with intersections increase algebraically with an exponent near 1. This behavior matches that observed in Ref. 7 and can be understood by recognizing that the distribution of distances to the closest object is well described by an exponential that is independent of the value of $\alpha$. As a result, the fraction of intersections is directly proportional to $d$ for $d \ll 1$, leading to a slope of 1 in the plot. We have not systematically investigated whether the algebraic trend changes in the limit where $d$ becomes comparable to the size of a voxel. Concerning the limit of small $d$, for $N = 6250$ in the case of Fig. 5e, the minimum overlap fraction we can detect is of the order of $(N\,(1 + \ln N))^{-1} = 1.6 \times 10^{-5}$.

Even though Fig. 5d,e shows that a naïve application of the model would not satisfy volume exclusion, these results provide guidance as to how to satisfy this constraint. First, the scaling of the number of edge crossings in Fig. 5d is relevant to understand the extent to which edge-edge interactions influence the stiffness of the material under strain. When the strain is zero, the material behavior derives from the stretching and bending of the edges. Once strained as in the figure, the number of edge crossings serves as a proxy for when contact interactions between edges begin to contribute non-negligibly to the overall behavior. Second, while Fig. 5e shows that intersections can occur at very small edge thicknesses, these intersections can be avoided in at least 4 ways: (i) deleting one edge involved in each overlap; (ii) creating nodes at edge-edge intersections and re-wiring the original edges to the nodes they intersect with; (iii) increasing the radii of the edges from zero only until they touch a nearby edge or node, allowing for edges of different thickness; and (iv) allowing the edges to be curved. We note that (i) can be augmented with rejection sampling used in random sequential deposition algorithms[8] and (iii) is amenable to generalization according to the Lubachevsky-Stillinger algorithm[9], which is used to generate packings of disks and spheres in other contexts.

## S2. Procedure to generate the images in Fig. 2

We use the Lubachevsky-Stillinger algorithm[9] to generate the disordered set of $N = 400$ spheres shown in Fig. 2a, where the principal volume consists of a cube of $S^3 = 1000$ voxels, with periodic boundary conditions. The spheres are initialized to be bi-disperse, with 64% having an initial radius of $d = 0.04$ and the remaining 36% having a radius $d = 0.07$ to prevent crystallization. The expansion rate of the spheres is set to 0.1, and particle collisions are governed by a Hertzian potential, which is proportional to $(d_i + d_j - \ell_{ij})^{2.5}$ when $(d_i + d_j - \ell_{ij}) > 0$, and zero otherwise, where $d_i$ and $d_j$ are the radii of spheres $i$ and $j$, and $\ell_{ij}$ is the Euclidean distance between their centers. The particles are allowed to grow until they become jammed. The granular network shown in Fig. 2a is the result of this process. The truss network in Fig. 2b consists of nodes placed at the center of each sphere and edges connecting spheres that are in contact in Fig. 2a. The plate network in Fig. 2c is created by computing the Voronoi ridges between the sphere centers and retaining the faces of the Voronoi ridges lying between spheres that are in contact in Fig. 2a. In this way, the topology and geometry of all three networks are captured by the same spatially embedded graph representation in Fig. 2d.

## S3. Supplementary figures

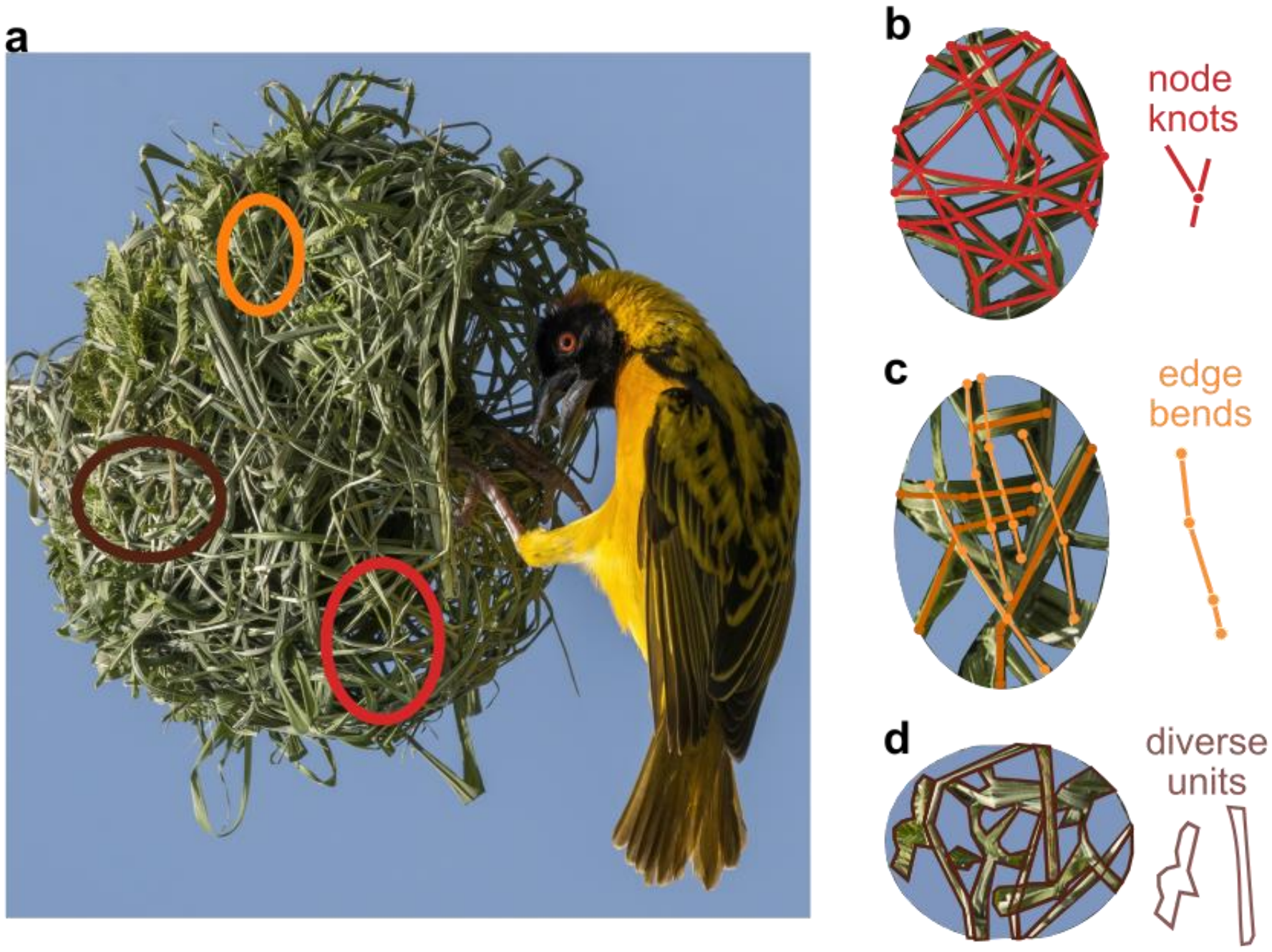

**Fig. S1: Topological, geometrical, and parametric irregularity illustrated in a bird's nest. a,** Nest of a black-headed weaver (image from Charles J. Sharp, CC BY-SA 4.0, via Wikimedia Commons, mirror-reflected and rotated for clarity). **b-d,** Magnified regions of the nest color coded as in (a) that illustrate knots between nest fibers (b), the bending of each fiber (c), and the different types of fibers (d) that form the nest structure.

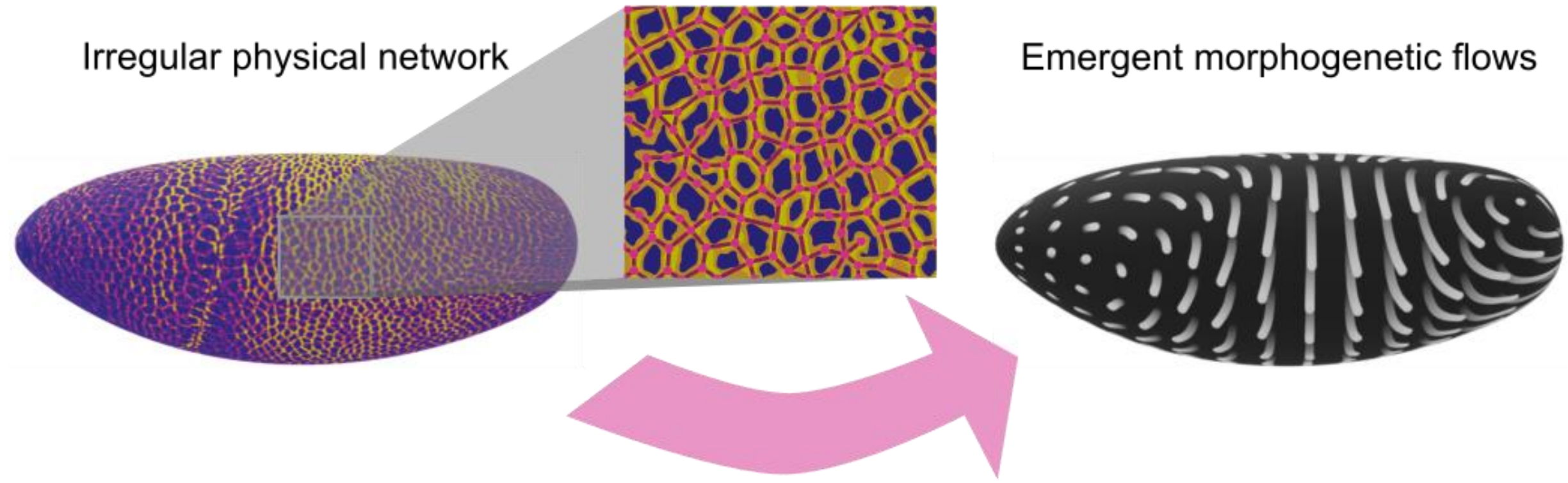

**Fig. S2: Natural physical learning in the myosin protein network in the fruit fly embryo.** The myosin protein network (left) learns, by responding to mechanical forces, where to place signaling molecules called morphogens (right) that direct the placement of cells forming the head and tail of the larva. The learning rules were inferred from microscopy data using interpretable machine learning. Image adapted from Ref. 10 with permission.

Supplementary references